\documentclass[twocolumn]{aastex701}

\usepackage{amsmath}
\usepackage{appendix}
\usepackage{mathtools}
\usepackage[caption=false]{subfig}
\usepackage{enumitem}
\usepackage{savesym}
\savesymbol{tablenum}
\usepackage{siunitx}
\restoresymbol{SIX}{tablenum}
\usepackage{xcolor}
\newcommand{\change}[1]{#1} 

\begin{document}

\title{Primary Beam Chromaticity in HIRAX: \\I. Characterization from Simulations and Power Spectrum Implications}

\author[0000-0002-0015-3870]{Ajith Sampath}
\affiliation{Astrophysics Research Centre, University of KwaZulu-Natal, Westville Campus, Durban 4041, South Africa}
\affiliation{School of Mathematics, Statistics and Computer Science, University of KwaZulu-Natal, Westville Campus, Durban 4041, South Africa}
\affiliation{Université de Genève, Département de Physique Théorique and Centre for Astroparticle Physics, 24 quai Ernest-Ansermet, CH1211 Genève 4, Switzerland}
\email{ajithsampath1997@gmail.com}

\author[0000-0003-1204-3035]{Devin Crichton}
\affiliation{Institute for Particle Physics and Astrophysics, ETH Zurich, 8092 Zurich, Switzerland}
\email{dcrichton@phys.ethz.ch}

\author[0000-0001-6606-7142]{Kavilan Moodley}
\affiliation{ Astrophysics Research Centre, University of KwaZulu-Natal, Westville Campus, Durban 4041, South Africa}
\affiliation{School of Mathematics, Statistics and Computer Science, University of KwaZulu-Natal, Westville Campus, Durban 4041, South Africa}
\email{Moodleyk41@ukzn.ac.za}

\author{H. Cynthia Chiang}
\affiliation{Department of Physics, McGill University, Montreal, Quebec H3A 2T8, Canada}
\email{hsin.chiang@mcgill.ca}

\author{Eloy De Lera Acedo}
\affiliation{Department of Physics, University of Cambridge, Cambridge, Cambridgeshire CB3 0HE	UK}
\email{ed330@cam.ac.uk}

\author{Simthembile Dlamini}
\affiliation{Department of Physics and Astronomy, University of the Western Cape, Robert Sobukwe Road, Bellville 7535, Cape Town 7535, South Africa. }
\email{simther4111@gmail.com }

\author{Sindhu Gaddam}
\affiliation{ Astrophysics Research Centre, University of KwaZulu-Natal, Westville Campus, Durban 4041, South Africa}
\affiliation{School of Mathematics, Statistics and Computer Science, University of KwaZulu-Natal, Westville Campus, Durban 4041, South Africa}
\email{gaddamsindhu20@gmail.com}

\author{Kit M. Gerodias}
\affiliation{Department of Physics, McGill University, Montreal, Quebec H3A 2T8, Canada}
\email{kit.gerodias@mail.mcgill.ca}

\author{Quentin	Gueuning}
\affiliation{Department of Physics, University of Cambridge, Cambridge, Cambridgeshire CB3 0HE	UK}
\email{qdg20@cam.ac.uk}

\author{N. Gupta}
\affiliation{Inter-University Centre for Astronomy and Astrophysics, Post Bag 4, Ganeshkhind, Pune 411 007, India}
\email{nrjgupta@gmail.com}

\author{Pascal Hitz}
\affiliation{Institute for Particle Physics and Astrophysics, ETH Zurich, 8092 Zurich, Switzerland}
\email{hitzpa@phys.ethz.ch}

\author{Aditya Krishna Karigiri Madhusudhan}
\affiliation{Department of Physics, McGill University, Montreal, Quebec H3A 2T8, Canada}
\email{aditya.karigirimadhusudhan@mail.mcgill.ca}

\author{Shreyam Parth Krishna}
\affiliation{LASTRO, Ecole Polytechnique Federale Lausanne, Chemin Pegasi 51, Versoix, Geneva 1290, Switzerland. }
\email{shreyam.krishna@epfl.ch}

\author{V. Mugundhan}
\affiliation{Department of Space, Planetary, Astronomical Sciences and Engineering, Indian Institute of Technology, Kanpur 208016, India.}
\email{mugundhanv@iitk.ac.in}

\author{Edwin Retana-Montenegro}
\affiliation{ Astrophysics Research Centre, University of KwaZulu-Natal, Westville Campus, Durban 4041, South Africa}
\affiliation{School of Mathematics, Statistics and Computer Science, University of KwaZulu-Natal, Westville Campus, Durban 4041, South Africa}
\email{edwinretana@gmail.com}

\author{Benjamin R.B. Saliwanchik}
\affiliation{Instrumentation Department, Brookhaven National Laboratory, Upton, New York, USA}
\email{bsaliwanc@bnl.gov}

\author{Mario G. Santos}
\affiliation{Department of Physics and Astronomy, University of the Western Cape, Robert Sobukwe Road, Bellville 7535, Cape Town 7535, South Africa. }
\email{mgrsantos@uwc.ac.za}

\author{Anthony Walters}
\affiliation{ Astrophysics Research Centre, University of KwaZulu-Natal, Westville Campus, Durban 4041, South Africa}
\affiliation{School of Mathematics, Statistics and Computer Science, University of KwaZulu-Natal, Westville Campus, Durban 4041, South Africa}
\email{waltersa@ukzn.ac.za}

\correspondingauthor{Ajith Sampath}
\email{ajithsampath1997@gmail.com}

\begin{abstract}

The Hydrogen Intensity and Real-time Analysis eXperiment (HIRAX) is an upcoming radio interferometric telescope designed to constrain dark energy through the 21cm intensity mapping of Baryon Acoustic Oscillations (BAO). Instrumental systematics must be controlled and carefully characterized to measure the 21cm power spectrum with fidelity and achieve high-precision constraints on the cosmological parameters. The chromaticity of the primary beam is one such complicated systematic, which can leak the power of spectrally smooth foregrounds beyond the ideal horizon limits due to the complex spatial and spectral structures of the sidelobes and the mainlobe. This paper studies the chromaticity of the HIRAX Stokes I primary beam and its effects on accurate measurements of the 21cm power spectrum. To investigate the effect of chromaticity in the 21cm power spectrum, we present a physically motivated beam modeling technique, which uses a flexible basis derived from traditional optics that can account for higher-order radial and azimuthal structures in the primary beam. We investigate the impact of imperfect knowledge of the mainlobe and sidelobes chromaticity in the power spectrum space by subtracting a simple foreground model in simulated snapshot visibilities to recover the H$\textsc{i}$ power spectrum. Additionally, we find that modeling up to the octupolar azimuthal order feature (fourth-order angular variation) in the primary beam is sufficient to reduce the leakage outside the wedge with minimal bias.

\end{abstract}


\section{Introduction}\label{sec:intro}
The current era of precision cosmology has led to next-generation experiments that refine the standard cosmological model by placing tighter constraints on cosmological parameters. Although these recent cosmological observations have improved our knowledge of the observable universe, many aspects still need to be fully understood and uncovered, such as the nature of dark energy. Dark energy is the unknown driving energy that causes the accelerated expansion of the universe, which makes up approximately three-fourths of the total content of our universe. By surveying large volumes of the spatial distribution of large-scale structures throughout the universe's history with high precision, one can infer the evolution of the dark energy equation of state parameters. This survey is possible by studying the chronological change of Baryonic Acoustic Oscillations (BAO), which are primordial sound waves in the early universe imprinted on the matter distribution with a current characteristic length scale of $\approx 150$ Mpc. This BAO feature is observed through the correlations imparted onto the large-scale structures traced by galaxies and neutral hydrogen (H\textsc{i}) \citep{Chang2008}. For example, the Sloan Digital Sky Survey (SDSS) measures BAOs using optical galaxies as tracers \citep{Blanton2017}. The sheer abundance of hydrogen and its optically thin nature provide a viable approach to make this extensive cosmological survey possible. 

The H\textsc{i} present from the recombination epoch to the present time serves as a promising tracer of these large-scale structures. The classically forbidden hyperfine spin-flip transition of H\textsc{i} emits a photon with a narrow line-width at 1420 $MHz$ ($21~cm$). Post the epoch of reionization (EoR), the remaining unionized H\textsc{i} was trapped in massive clouds and shielded from the ionizing ultraviolet photons, now called Lyman-$\alpha$ absorbers. Therefore, it is possible to map this trapped H\textsc{i} signal spatially along with the redshift information to create a 3D H\textsc{i} field. However, this signal is very weak in amplitude, and the current telescopes are systematics-limited. The modern interferometers use the well-known intensity mapping (IM) technique to map this 21cm signal at a low resolution \citep{Bull2015}, i.e., mapping unresolved galaxies combined in a single pixel, boosting the signal amplitude.\\

The Hydrogen Intensity and Real-time Analysis eXperiment (HIRAX) \citep{Newburgh2016, Crichton2022} is a radio interferometer array currently in development to be deployed at the SARAO site in the Karoo in South Africa with an initial 256-element array. The primary goal of the telescope is to survey the southern sky through low-resolution intensity mapping of the redshifted H\textsc{i} 21cm line and thereby to map out the distribution of the large-scale structure over a redshift range of 0.775 $<$ z $<$ 2.55 which corresponds to an observing frequency range of 400 -- 800 $MHz$. The survey covers an angular region over $\sim$15,000 square degrees of the southern sky. This survey is statistically robust enough to produce $\approx$7 percent constraints on the dark energy equation of state parameters when combined with \textit{Planck} \citep{planck2021} measurements. HIRAX will complement CHIME \citep{Bandura2014,chime2022} and CHORD \citep{chordwhitepaper} in Canada, surveying the Northern sky and HIRAX covering a declination range of $-60^\circ \leq \delta \leq 0^\circ $ in the Southern sky, and overlapping with a range of other extensive surveys such as eBOSS \citep{Alam2021}, DESI \citep{desi}, LSST/Rubin \citep{Ivezić_2019} and Euclid \citep{Amendola2018} and other ground-based Cosmic Microwave Background (CMB) surveys, opening a wide-window of potential cross-correlation studies \citep{Crichton2022}.\\

Since the foregrounds (both galactic and extra-galactic origin) are up to six orders of magnitude brighter in power than the signal of interest, even after rigorous cleaning of foregrounds, the residuals would still contaminate the power spectrum when the systematics are not controlled. One significant systematic that causes foreground leakage is the chromaticity of the beam, i.e., the frequency dependence of the primary beam. The reason behind these residuals, even after foreground cleaning, is the mis-modeling of the primary beam. The primary beam is the response pattern of an individual antenna in a radio telescope array. This mis-modeling will also make the foreground-free signal window smaller, as later discussed in this work. \\

The aim of this study is to model the full beam of any radio telescope beyond the first null analytically and investigate the effects of sidelobes in cosmological measurements. Some previous studies tried and succeeded in creating a functional model with up to two or three sidelobes \citep{Smirnov2018, Iheanetu2019, Asad2021, Young2013, BuiVan2017}. However, even with those models, the signal contamination due to strong sources residing in farther sidelobes can be significant. Also, the true flux scales are traditionally calibrated with a beam model. \citealp{Gan2022} shows that including a beam model during the direction-dependent calibration process significantly improves its overall performance. Our beam modeling formalism provides a full beam model vital for solving antenna beam-based errors in absolute flux calibration. Understanding the realistic behavior of beams is essential to achieve high SNR during calibration and foreground cleaning in 21cm intensity mapping surveys. Although the previous works on beam modeling have some applications in imaging, no specific formalism helps understand the beams extending far into their sidelobes and the sidelobe chromaticity in 21cm intensity mapping and EoR experiments. 

This paper is the first in a two-part series addressing primary beam chromaticity in HIRAX. In this first part, we focus on modeling and characterization of the HIRAX primary beam and pointing out its implications for the power spectrum estimation. The second part will build on these foundations by applying the modeling framework to real drone-mapped beam data and presenting foreground mitigation strategies. 

The paper is structured as follows. In section \ref{sec:fg}, we provide an overview of current challenges in analyzing the statistical detection of H\textsc{i} signal with 21cm intensity mapping experiments due to systematics by emphasizing the significance of beam modeling beyond the mainlobe. The fitting results of the mainlobe of the primary beam and its chromaticity are discussed in detail in section \ref{sec:fitsim}, which includes some basic introduction to the electromagnetic (EM) simulated beams. In section \ref{sec:ps}, we validate our formalism and investigate the leakage of foregrounds beyond ideal limits due to beam modeling uncertainties caused by neglecting sidelobes using simulated visibilities. Sections \ref{sec:disc} and \ref{sec:conc} contain further discussions and conclusions along with the future scope of this work, respectively. 
In appendix \ref{sec:zern}, we describe the mathematical formalism of our basis function and the motivation to choose such a basis, and in appendix \ref{sec:ps_form}, we revisit the measurement equation and formulate the power spectrum through the delay approach. This work introduces the analytical basis and studies the chromaticity of the simulated HIRAX Stokes I primary beam's mainlobe and sidelobes. We also investigate and quantify the foreground leakage due to sidelobe modeling errors and provide insights on how far one can and should model the primary beams. 

\section{Foregrounds and beam chromaticity}\label{sec:fg}
Like any other 21cm experiment looking for a faint background signal of cosmological significance, HIRAX also suffers from the well-known foreground contamination problem, with foregrounds several orders of magnitude brighter than the background signal dominating the power spectrum \citep{Pritchard2012}. The next-generation EoR experiments such as Hydrogen Epoch of Reionization Array(HERA) \citep{DeBoer2017}, and Square Kilometer Array-low(SKA-low) \citep{Labate2017}, as well as post-EoR intensity-mapping experiments such as HIRAX, and MeerKAT, can, however, make detection by using an approach called ``foreground avoidance," where one does not have to subtract the foregrounds from the data, but rather avoid the region of contamination and access the clean window in Fourier space. In any interferometric observations, the foregrounds are generally smooth in frequency and are confined within a wedge-like region in cylindrical Fourier space ($k_{\parallel},k_{\bot}$: Fourier modes parallel and perpendicular to the line of sight) \citep{Datta2010, Morales2012, Vedantham2012}. This characteristic spectral behavior of foregrounds is critical to wide-field interferometers since one can exploit this smooth spectral behavior to avoid or subtract them from the dataset. The clean 21cm signal window is at higher $k_{\parallel}$ and lower $k_{\bot}$ regions, as observed and confirmed by EoR experiments such as PAPER \citep{Parsons2014} and MWA \citep{Tingay2013, Kolopanis2023}. 

However, there are two disadvantages to this approach. One is the leakage beyond the horizon limit (the theoretical limit for smooth spectrum foregrounds) due to the interaction between the chromatic response of the interferometer and the otherwise spectrally smooth foreground emission, which can cause these confined foregrounds to ``leak" into the clean signal window. This inherent chromaticity of the interferometer allows the foregrounds to spill into higher $k_{\parallel}$ modes. Using foreground simulations, \citealp{Thyagarajan2016} demonstrated that for contamination in every delay mode by the antenna gains, foreground power contaminates the signal with several orders of magnitude in power. \citealp{Pober2016} established the importance of addressing the sidelobe chromaticity in analysis and emphasized the need for wide-field foreground removal techniques. \citealp{Josaitis2022} showed that cross-coupling features in the primary beam can also leak foregrounds outside the wedge, irrespective of time, baseline length, and orientation. Another downside to the foreground avoidance approach is the loss of sensitivity in the signal since the lower $k$-modes, especially the BAO scales we are interested in $k \sim 0.05$ to $0.15 ~ Mpc^{-1}$, which correspond to the large scales, having higher signal-to-noise ratio but entirely dominated by foregrounds \citep{Chapman2016, Kim2022}. 

An alternative approach is foreground cleaning (or subtraction), where, in the case of a non-blind cleaning algorithm, the spectral smoothness of foregrounds is exploited to subtract the foregrounds from the dataset by fitting a low-order polynomial \citep{Bowman2009, Liu2009}. In blind foreground cleaning algorithms, the standard Principal Component Analysis (PCA) \citep{Alonso2014, BigotSazy2015} is used.  There are some semi-blind algorithms, for example, Karhunen-Loeve transforms \citep{Shaw2014}, which are found to be helpful. However, only a non-blind, well-informed algorithm can cope with systematic effects and improve foreground subtraction accuracy. In contrast, blind algorithms can lead to severe signal loss, bias, and spurious artifacts in the cleaned data \citep{Spinelli2021}. More recently, modern machine learning methods (deep learning) have been used in foreground cleaning \citep{Makinen2021, Ni2022}. Since the chromaticity of the antenna beam response can become entangled with the smooth foregrounds, usually referred to as ``mode-mixing" \citep{Morales2012}, this leads to unsatisfactory foreground subtraction. \change{The poor foreground subtraction} is generally dominated by incorrectly subtracted sources towards the horizon, and this introduces a bias in the 2D power spectrum of up to two orders of magnitude higher than the expected signal for EoR experiments \citep{Joseph2019}. For example, CHIME found that foregrounds can be effectively cleaned down to $ k_{\parallel} \sim 0.02 h^{-1} Mpc$ well within the foreground wedge, even with polarization, given perfect knowledge of the instrument and unbiased power spectrum can be recovered if the per-feed beam width is measured to 0.1$\%$ accuracy \citep{Shaw2014}. Also, for foreground subtraction to be precise, accurate wide-field primary beam calibration is needed to subtract the bright foregrounds falling in far-sidelobes, and they should be treated as a wide-field contaminant \citep{Pober2016}. So, a complete characterization of the primary beam response is needed for any foreground cleaning procedure that aims to achieve high precision H\textsc{i} signal recovery.

\section{Fitting simulated beams}\label{sec:fitsim}

Our primary goal is to understand the impact sidelobe chromaticity has on measurements of the 21cm power spectrum for HIRAX, and it serves as the primary motivation for opting for an analytical representation in image space, as it simplifies the application of this method to cosmological simulations. HIRAX antennas have a circular aperture of \SI{6}{\meter} diameter and a dual-polarized feed supported by four feed legs \citep{Saliwanchik2021}. The beam response depends heavily on this configuration, the dish surface deformities, and the operating frequency. \citealp{Xiang2019} studied the dish surface errors by modeling the aperture using Zernike polynomials (ZP hereafter). A recent study \citep{Sekhar2022} modeled the aperture illumination function (AIF) using the conventional Zernike polynomials over the aperture space and applied the modeling on Very Large Array (VLA), MeerKAT, and Atacama Large Millimeter Array (ALMA) measurements. \change{This is in fact the same model, but in the aperture plane. We opt for image space modeling, which offers direct control over beamwidth, frequency scaling, and pointing errors, enabling targeted correction of beam-induced systematics.} There have also been similar studies on the image space recently by \citealp{Iheanetu2019} and \citealp{Asad2021} on VLA and MeerKAT beams using both principal component analysis (PCA) and Zernike basis. Their choice of analytical basis was conventional Zernike polynomials, and their goal was to find a sparse beam representation. We also want a sparse representation of our beam to be easily employed in the calibration pipeline. The modeling approaches mentioned above were applied to imaging telescopes to get the best flux estimation of the targeted source falling on the mainlobe. However, our modeling approach focuses on characterizing chromatic sidelobes in wide-field interferometers. 

\subsection{Electromagnetic simulation setup}\label{sec:emsims}
We use electromagnetic (EM) simulated power beams with no noise to understand the chromatic effects and test our formalism. \change{While the electric field beam (the elements of the Jones matrix) is a more fundamental quantity for describing polarization properties, here we focus on the more practically measurable power beam using forthcoming drone-based measurements. A similar basis has been applied to model the complex Jones matrix in \citealp{Wilensky2024}.} The electromagnetic simulations are carried out in the Computer Simulation Technology (CST) studio suite\footnote{CST Studio Suite: \url{https://www.3ds.com/products-services/simulia/products/cst-studio-suite}}. In our simulations, the model is simplified to a dish-and-feed configuration where the four feed legs and the cables are removed to reduce the computational cost. The final HIRAX dish design pushes the feed deeper into the dish to avoid cross-talk between other antennas and ground-spill, making the effective diameter of the dish $\sim 4.5~m$ and f-number $f/0.21$. In this setup, the feed hovers $126~cm$ above the vertex of the parabolic dish. This simple setup is not the ground truth since, in reality, we have legs to support the feed and cables for power supply and data collection running, thus blocking the aperture. But, previous CST simulations that did include the feed legs and cabling showed they had minimal impact on the beams, especially if the legs were made of non-conductive materials and the cabling was aligned along the boresight \citep{Saliwanchik2021}. Also, we are only considering the radiation pattern of an isolated dish here. However, in reality, the antennas in the closely packed HIRAX array can perturb the radiation patterns of each other \citep{Gueuning2022}. \change{We find that the 2D radiation patterns are axially symmetric along both E-plane and H-plane, but azimuthally asymmetric, } even though the dish-feed setup is elementary. The radiation pattern is determined by the physical structure of the active element, i.e., the cloverleaf-shaped dual-polarized dipole antenna in our case. \change{The chosen shape of the feed active element induces more asymmetric features at high frequencies, and this simulated beam is expected to be representative of the finalized design to be deployed.} See table \ref{table1} for details of simulation parameters. In our procedure, we limit our spatial modeling of the beam to a radial distance of $75^\circ$ from the center. This limitation is imposed for faster computations, as our beam exhibits no significant features beyond this angular distance. \change{In fact, the percentage fraction of the total encircled power within this $75^\circ$ radius compared to the total beam is around $99.6\%$ in all frequencies.} Also, this method is flexible to model further sidelobes in our fitting routine if needed.

\begin{table}
\begin{tabular}{c|c}
\hline
\hline
\multicolumn{2}{c}{CST simulation parameters}\\
\hline
\hline
Configuration & dish and feed only \\
Dish diameter & \SI{6}{\meter} \\
Focal ratio & $f$/0.21 \\
Bandwidth & 400 - 800 $MHz$ \\
$N_{channels}$ & 81 \\
Spectral resolution & 5 $MHz$\\
\hline
\end{tabular}
\caption{ Parameters used for simulation in CST studio. Actual dish diameter, focal ratio, and bandwidth are used, but we used only 81 channels here to reduce the computation time for the fitting routine described in the following sections.}
\label{table1}
\end{table}

\subsection{Gaussian fit}\label{sec:gfit}
\begin{figure}
    \centering
    \includegraphics[width=9cm]{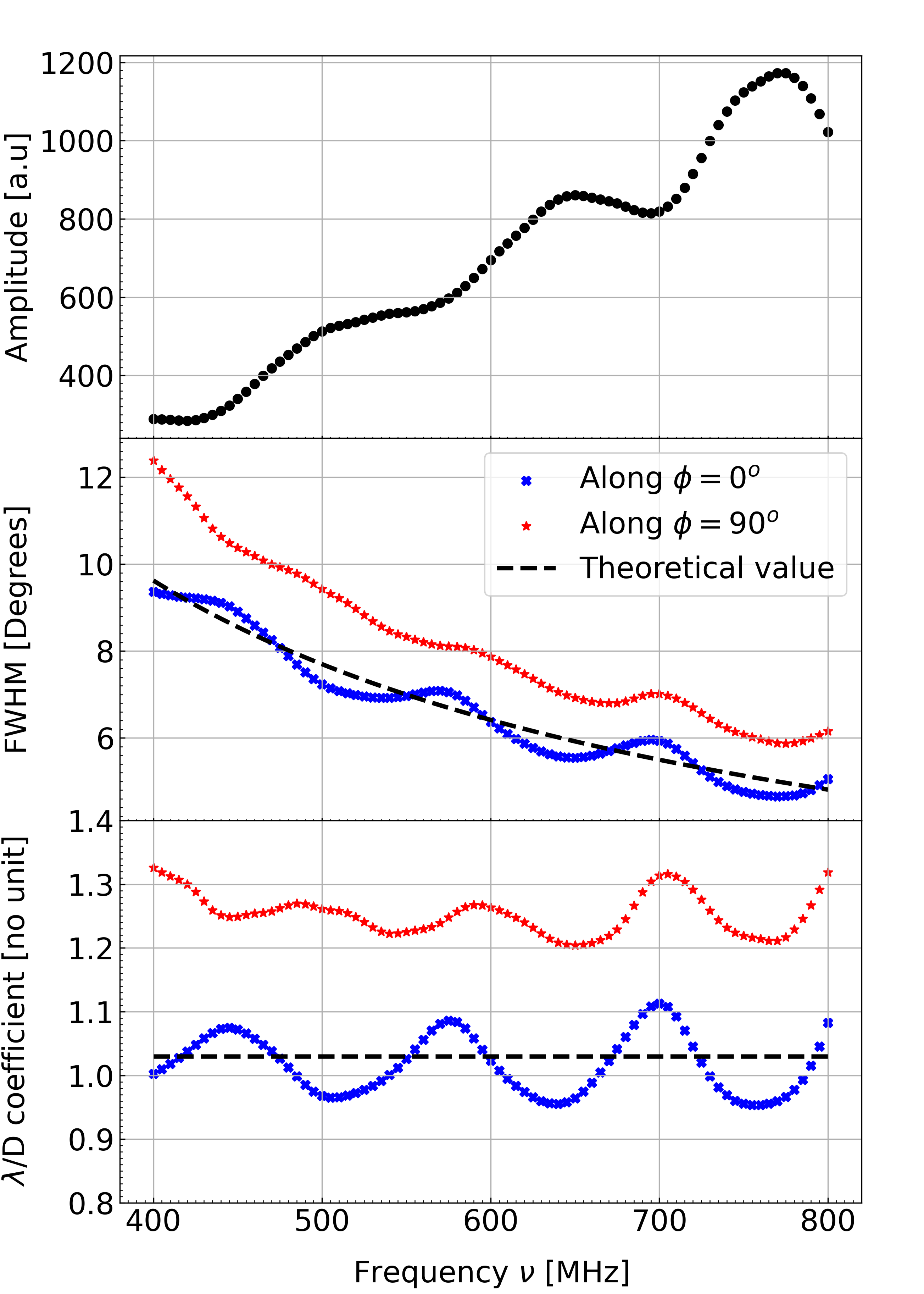}
    \caption{Parameters estimated from the elliptical Gaussian fit to the CST simulated Stokes I beam. The x-axis corresponds to the frequency in $MHz$. The y-axis represents \textbf{\textit{Top:}} the peak amplitude in arbitrary units, \textbf{\textit{Middle:}} full width at half maximum in degrees, and  \textbf{\textit{Bottom:}} $\lambda / D$ coefficient, respectively. The dashed line in the middle and bottom plots represents the theoretical FWHM $\theta_t$. } %
    \label{fig:1}
\end{figure}

We first start with fitting the mainlobe with an elliptical 2D Gaussian function to calculate the beamwidth. We use the \textit{Nelder-Mead} minimization algorithm \citep{Nelder1965} to find the optimal values of peak amplitude and beamwidth (\textbf{$\sigma_x$},\textbf{$\sigma_y$}) along $x$ and $y$-axis (semi-major and semi-minor axis). We do this to all 81 beam maps corresponding to the 400 - 800 $MHz$ frequency range. The Gaussian fit captures the change in amplitude of beams with frequency and the change in full width at half maximum (FWHM), revealing the mainlobe chromaticity. Figure \ref{fig:1} shows the gain variation with frequency and the spectral structure in the mainlobe. The beamwidth calculated from the Gaussian profile fits for all channels deviates significantly from the theoretical beamwidth $ \theta_t \sim 1.03 \lambda/D_{eff}$ (where $\lambda$ denotes the wavelength and $D_{eff} \approx 4.5 ~ m$ denotes the effective diameter of HIRAX dish). The $\lambda/D$ coefficient of the fitted Gaussian profile shows a sinusoidal pattern. This pattern is called ``beam-ripple," which was observed and well studied in MeerKAT \citep{Matshawule2021} and is due to the interaction between primary and secondary reflectors \citep{deVilliers2013}. In HIRAX, this characteristic ripple is due to the reflections between the feed and the dish surface. This ripple corresponds to an average length scale of $2.5m$, approximately twice the focal length ($1.26m$). This ripple will be very significant in foreground cleaning as it adds structure to the chromatic response of the instrument, which mixes with the otherwise spectrally smooth foregrounds. The ripple may become more complicated in reality with more aperture blockage, cross-coupling between antennas, and variations in time. Spectral structures like this can also originate from signal chains. For example, the reflections within the signal path and the instrumental signal chain can add spectral structure to the correlated visibility data. Longer delays in the signal path can taint the finer frequency scales and can, therefore, cause leakage in the power spectrum \citep{Beardsley2016, EwallWice2016}. 

\subsection{Zernike Transform fit}\label{sec:zfit}
\begin{figure*}[ht!]
    \centering
   \includegraphics[width=17.5cm]{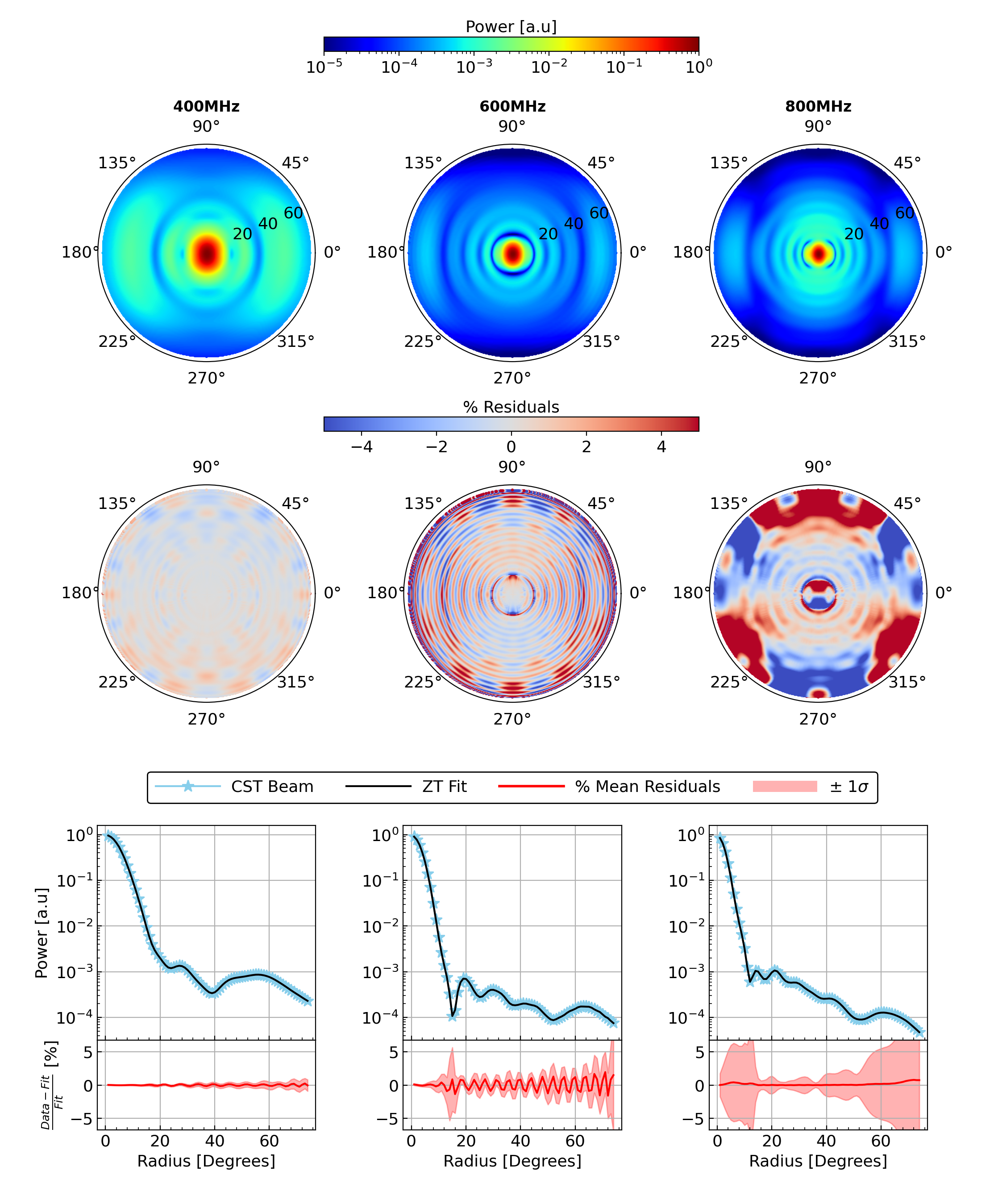}
    \caption{\textbf{\textit{Top:}} ZT Model fitted to the simulated CST Stokes I beam, \textbf{\textit{Middle:}} Residual difference between CST beam and model in linear scale in percentage units of with respect to the model, and \textbf{\textit{Bottom:}} Radial profiles at 400,600 and 800 $MHz$ (each column from left to right respectively) with simulated beam in \textit{blue}, fitted model in \textit{black}, mean percentage residuals in \textit{red} and $1\sigma$ standard deviation in \textit{shaded red}. \\} %
    \label{fig:2}
\end{figure*}

\begin{figure*}
\centering 
\includegraphics[width=8.5cm]{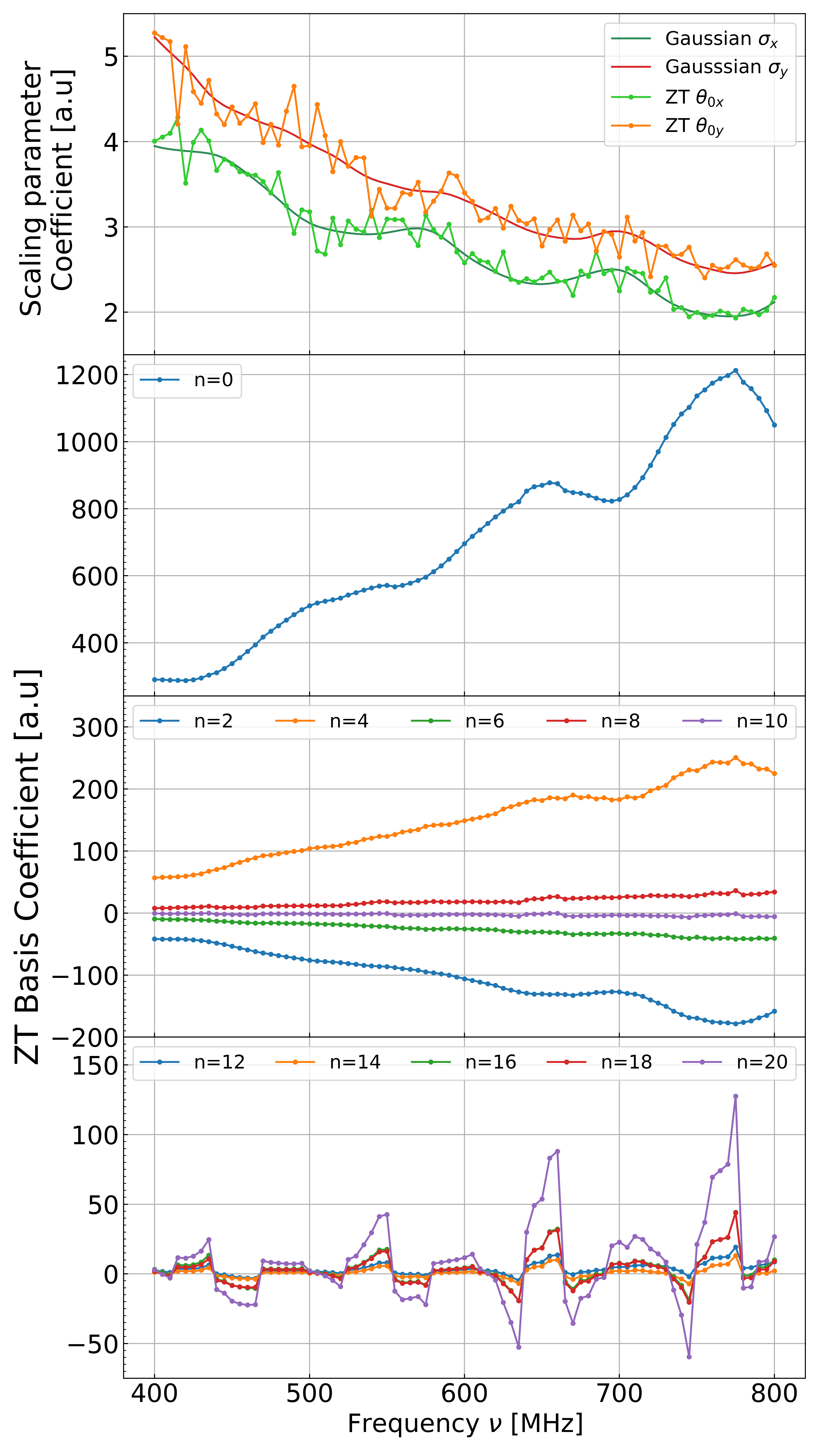}
\qquad
\includegraphics[width=8.5cm]{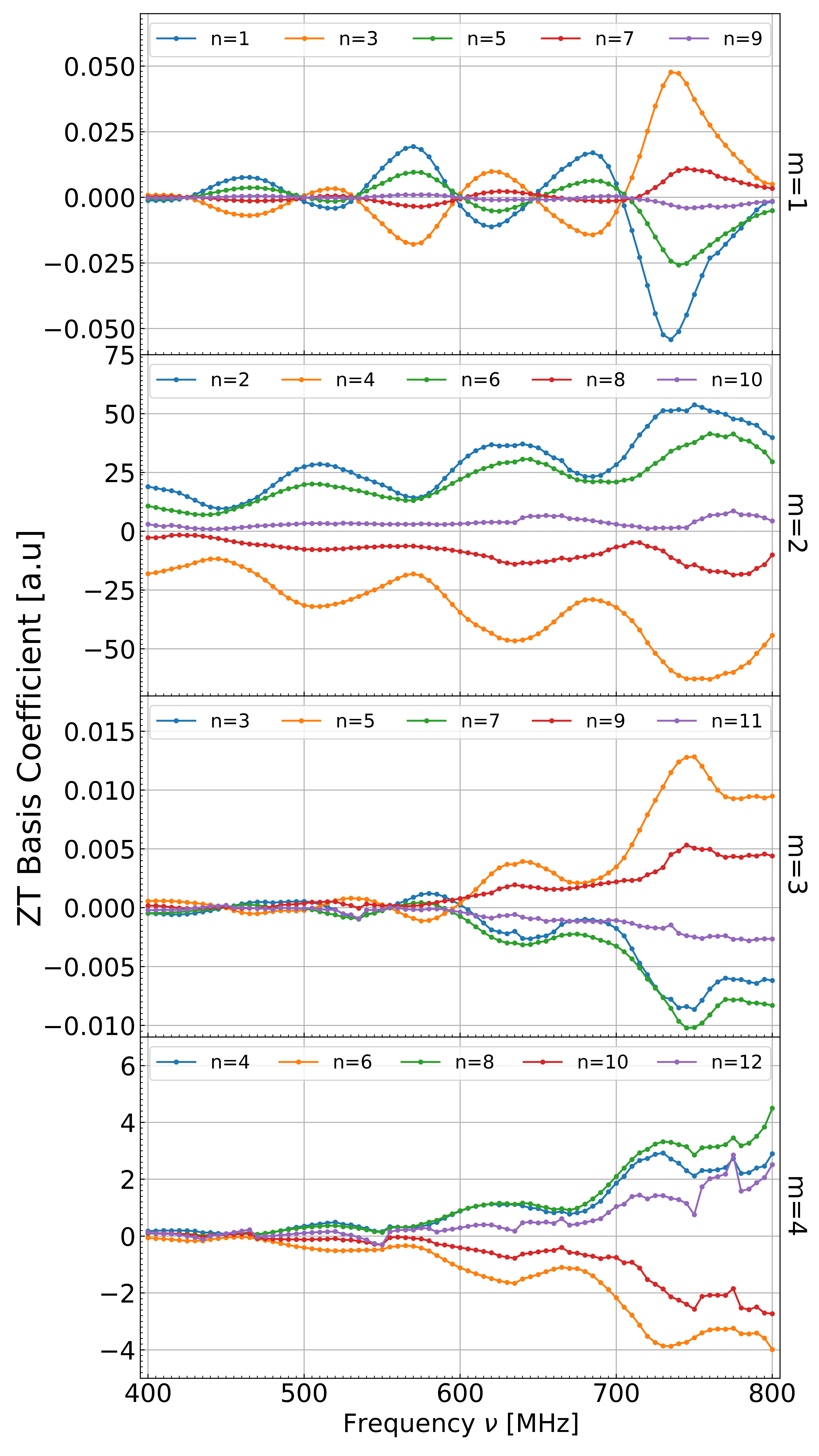}
\caption{The fitted coefficient energies as a function of frequency in $MHz$. \textit{\textbf{Left column:}} \change{\textit{Top:} Scaling parameter as a function of frequency. Comparison of Gaussian mainlobe fit scaling parameters $(\sigma_x,\sigma_y)$ and ZT basis fit scaling parameters $(\theta_x,\theta_y)$. \textit{Bottom three:} Chromaticity of the first few radial (m=0) modes. \textit{\textbf{Right column:}} Chromaticity of the first few non-radial modes. \\}}
\label{fig:3}
\end{figure*}

Although the simple Gaussian model captures the chromaticity of the mainlobe where the foregrounds are highly sensitive, it is not sufficient for precision cosmology, especially for wide-field interferometers, since sidelobes can be more problematic, mainly when we try not to avoid foregrounds but rather clean them down to the lower $k_{\parallel}$ modes. Therefore, we opt for a beam modeling procedure where one can fully model the primary beam to one's requirement. Our choice of basis here is the analytical form of the Fourier transform of the ZP. We call them Zernike Transforms (ZT hereafter) and are given by 
\begin{equation}
\label{eq:1}
    \mathcal{Z}^m_n = \frac{e^{im\phi}}{2\pi i^m}~ (-1)^{(n-m)/2} ~ \frac{J_{n+1}(\theta/\theta_0)}{\theta/\theta_0} 
\end{equation}
and using this basis, we can model our beams such that our reconstructed beam is defined as,
\begin{equation}
\label{eq:2}
    \mathcal{B} (\theta,\phi) = \sum_{n,m} \alpha^m_n ~ \mathcal{Z}^m_n(\theta,\phi) 
\end{equation}
where $\alpha^m_n ~ \in \mathbb{C}$. \change{Here, $\mathcal{B}$ denotes the unpolarized power beam of a single antenna, defined as the modulus squared of its complex electric field pattern $\mathcal{B} = |\mathcal{B}_i|^2$, as shown in Appendix \ref{sec:ps_form}, Equation \ref{eq:b10}.} Also, $\theta_0$ here is a frequency-dependent scaling parameter introduced to facilitate fitting, and is not an intrinsic part of the basis functions.
We describe the mathematical formalism behind this in detail in Appendix \ref{sec:zern}. Our basis has a significant advantage over other analytical formalisms in terms of wide-field modeling because, as the basis order rises, the first peak shifts further from the center, reflecting the spacing of the zeros of the Bessel functions of the first kind (see Equation \ref{eq:1} and Figure \ref{fig:1} in Appendix A). This unique feature can capture the farther sidelobes by increasing the order, enabling us to study their behavior separately. 

We fit the simulated CST beam with the ZTs using a numerical optimization routine (since this method will be applied to real drone-mapped data in the future) up to $75^\circ$ in angular distance from the beam center. We fit the 2D beam maps channel-wise, introducing a scaling parameter $\theta_0$ for the basis function. \change{Because the aperture-plane variable $r$ (in Equation \ref{eq:a1}) corresponds to a physical length, a natural wavenumber dependence arises in the Fourier transform. In our approach, this scaling is not imposed; instead, $\theta_0$ is fitted separately at each frequency, allowing the model to adjust freely to match the simulation.} Therefore, the fitting routine needs two input parameters,i.e., $\theta_0$, the scaling parameter which decides the extent of the basis, and $n_{max}$, the maximum number of bases being used.  We use the 2D Gaussian $\sigma$ from the Gaussian fit in the previous section (Section \ref{sec:gfit}) as the initial guess for $\theta_0$ for our optimization routine here, and we set the $n_{max}~=~100$. Although we are aware that this basis (Equation \ref{eq:1}) captures the aberration and effects in aperture space (for example, Coma and Astigmatism), we choose a subset of basis functions from this complete set. After thorough analysis and many fitting routines, we found that the complex part of the full basis contributes negligibly little to the overall fit, so we chose only the real part for further fitting. We also mask the negative order azimuthal ($m$) modes ($az$-modes hereafter) since the real parts of the negative and positive $az$-modes are identical (as shown in the left panel of Figure \ref{fig:8}), which causes degeneracy in the fitting procedure. So, we use the chosen reduced (subset) basis, which is just the real part of non-negative $az$-modes $Re\{\mathcal{Z}^m_n (n,m\geq0)\}$ (note that we did not remove or mask the modes where $m=0$; they are very critical to model our primary beam). So, now our original basis $\mathcal{Z}^m_n(\theta,\phi)$ in Equation \ref{eq:1} has changed to:
\begin{equation}
\label{eq:3}
    X_n^m (\theta,\phi) = \mathcal{R}e \Biggl\{ A_n ~ \frac{e^{i|m|\phi}}{2\pi i^{|m|}}
~~ (-1)^{\frac{n-|m|}{2}} ~~ \frac{J_{n+1}(\theta/\theta_0)}{(\theta/\theta_0)}  \Biggl\} \\
\end{equation}
\\
and the equation to solve for had changed from Equation \ref{eq:2} to 
\begin{equation}
    \label{eq:4}
    \mathcal{B} (\theta,\phi) = \sum_{n,m} \alpha^m_n ~X_n^m(\theta,\phi)
\end{equation}

where $\alpha^m_n ~ \in \mathbb{R}$ and $A_n$ is the normalization constant for the relation in Equation \ref{eq:a5}. \change{Here, Equations \ref{eq:3} and \ref{eq:4} correspond to a real-valued beam $\mathcal{B}$, where $\alpha_n^{-m} = \alpha_n^{m*}$ ensures cancellation of imaginary terms, and only positive-$m$ terms with real coefficients are retained. }

\begin{figure*}
\centering
    \includegraphics[width=18cm]{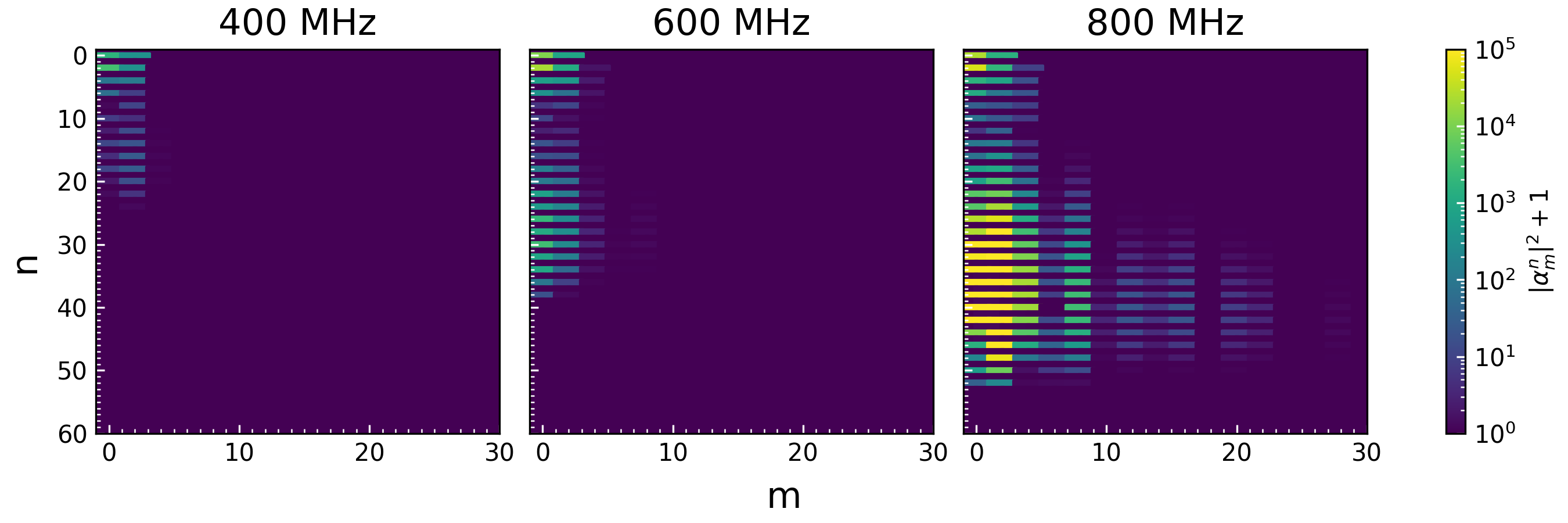} 
    \caption{Energy spectrum of the coefficients $\alpha_n^m$ at three different frequencies. A value of 1 is added to every coefficient to help with visualization. The power in each pixel of this energy spectrum denotes the significance of that particular ZT mode. The radial orders ($n$) are along the $y$-axis and azimuthal orders ($m$) along the $x$-axis \\}
    \label{fig:4}
\end{figure*}

Using weighted least squares, we up-weight the sidelobes by setting the weights $w_pix$ as the reciprocal of a chosen fraction of the beam amplitude, preventing overfitting to the mainlobe. This standard deviation is applied in the coefficient calculation, not as noise in the maps, but for the least squares and $\chi^2$ calculation as:
\begin{equation}
\label{eq:5}
    \sigma_{pix} \propto \mathcal{B}_{pix}.
\end{equation}
where $\mathcal{B}_i$ is the beam amplitude, and $pix$ is the pixel index. Since our simulations have no realistic noise, this assumption of $\sigma_i$ is idealistic but optimal to fit sidelobes. \change{In practice, $\sigma_{pix}$ will correspond to measurement errors associated with the drone beam maps \citep{Chang2015, tyndall}, and these measurement errors will be more in higher radii, but combining this $\sigma$ with measurement errors is left for future work. We tweak this to make the fitting work better, which would ensure that the fractional difference between the data and the model remains smaller even when the beam amplitude decreases.} The choice of this number can affect the fitting significantly since this gives the spatial weighting for the beam during the fitting process, and this weighting also helps up-weight the relatively undersampled high radii areas compared to oversampled low radii areas. We determine the goodness-of-fit using the reduced chi-square ($\chi_r^2$) method, which is given by
\begin{equation}
    \chi_r^2 = \frac{\chi^2}{k} 
\end{equation}
where
\begin{equation}
    \chi^2 = \sum \frac{(\mathcal{D}_i - \mathcal{B}(\theta_i,\phi_i))^2}{\sigma_i^2}  ~ ~ , ~ ~ ~ ~ k = N - M 
\end{equation}
with variance $\sigma _{i}^{2}$ calculated previously, data (CST beam) $\mathcal{D}$ , and model $\mathcal{B}(\theta_i,\phi_i)$. The degree of freedom $k$  equals the number of observations N minus the number of fitted parameters M. We decide the fit is reasonable when $\chi_r^2 \approx 1$. The fit coefficients are calculated by solving,

\begin{equation}
\label{eq:6}
    \hat\alpha = (X^T~X)^{-1}~X^T~\mathcal{B}.
\end{equation}

Figure \ref{fig:2} shows the model, residuals, and radial profile of the simulated beam, fitted model, and percentage residuals at 400 $MHz$, 600 $MHz$, and 800 $MHz$. The fitted model accurately captures the complex structures in the sidelobes of the beam, demonstrating the ability of the model to adapt to detailed spatial variations. The residual plot, showing the percentage differences between the simulated beam and the ZT model, indicates that while the overall structure is well captured, some dipole-like components remain unmodeled in higher frequencies, showing the complexity of the HIRAX beam and establishing a need for potential refinement in the model. Additionally, the residuals show that the model is accurate up to a percent level in most of the band and within a few percent at the highest frequency. The radial profile plot, which compares the CST beam and the fit, clearly indicates the model's overall performance. It also confirms that the ZT model effectively captures the beam's essential characteristics, establishing the accuracy of the fit.

To investigate the implications of chromatic sidelobes, we look at the frequency dependence of each ZT mode after fitting the CST beam (see Figure \ref{fig:3}). Since the sign and amplitude of the fit coefficients vary greatly, we plot the coefficients vs. frequency for each ZT mode and infer the chromaticity of the mainlobe and sidelobes separately. \change{We also compare the Gaussian mainlobe fit ($\sigma_x,\sigma_y$) with the ZT scaling parameter ($\theta_{0x},\theta_{0y}$). $\theta_{0x}$ and $\theta_{0y}$ show some fluctuations despite their overall agreement with the Gaussian $\sigma$ (top panel in left column of Figure \ref{fig:3}), since we set this as a free parameter in the fitting routine with no bounds.} The chromaticity of the first mode ($n=0,m=0$) as seen in the left column 2nd panel of Figure \ref{fig:3} resembles the top panel in Figure \ref{fig:1} since that mode captures the mainlobe structure (top panel in the left-sided plot of Figure \ref{fig:3}). Also, with increasing $n$ (with $m$ fixed at 0), we see the strength of the higher $rad$-modes in the higher end of the band growing. These radial $n$-modes ($rad$-modes hereafter) correspond to the sidelobes farther from the beam center, translating to the complexity of the HIRAX primary beam's sidelobes, especially at higher frequencies. \change{It is worth noting that the bottom panel of the left column in Figure \ref{fig:3} shows a lot of non-smooth chromatic structures in $n>10$ cases. But this could potentially be some numerical errors in the simulations, as we can notice a small jump in the $n=0$ case (second panel in left column) and in $2\geq n \geq 10$ case, especially around 650 $MHz$ and between 750 and 800 $MHz$. This should be investigated further in detail with various CST-based simulations by changing solvers in the simulation setup, but we defer this for future work.}

The complex chromaticity of the HIRAX beam is not just valid for the radial features ($m=0$) but also for the azimuthal features ($m\geq0$). Our modeling procedure can study the chromaticity of these non-radial modes ($az$-modes) separately, as seen in the right-sided plot of Figure \ref{fig:3}. The key takeaway from this figure is that even-numbered $az$-modes ($m=2, m=4$) contribute significantly to the overall fit. The structures also show an oscillating feature similar to the mainlobe chromatic structure, as seen in the (b) and (c) plots of Figure \ref{fig:1}.  

The initial fits require many basis functions since farther sidelobes require higher-order modes. However, once we fit the simulated beam, we reduce the number of modes used by picking the dominant modes and reconstructing the beam again to have a sparse representation. We define the contribution of ZT modes to the overall fit as the coefficient energy spectrum, which is the square of fitted coefficients as a function of $n$ and $m$ ($|\alpha_n^m|^2$). We plot the coefficient energy spectrum in Figure \ref{fig:4}. The first few $az$-modes are only required for the fit, which emphasizes that our HIRAX beam has low-order asymmetric features at the lower end of the band and more $az$-modes at the higher end. However, we also found that ranking these fitted modes based on the coefficient energy shows that very few modes contribute significantly to the overall fit. Nevertheless, as discussed later, these azimuthal asymmetries should be addressed since non-modeling or its omission can propagate errors further into the analysis. This characteristic azimuthal asymmetry is present in our beam due to the asymmetric illumination pattern of our dish. However, we will also see later that these low-order azimuthally asymmetric effects are not so complicated, as we may only need a few $az$-modes to represent the beam effectively. 

\section{Beam chromaticity in power spectrum}\label{sec:ps}
\subsection{Simulation procedure}\label{sec:procedure}

We use a 2-dimensional cylindrical power spectrum (2DPS) and a 1D power spectrum (1DPS obtained by radially averaging 2DPS) to investigate the modeling errors due to beam chromaticity. In this section, we will explain the procedure for simulating the visibilities for a realistic HIRAX array under the assumption of a complete sky model, i.e., we have perfect knowledge of the sky. Also, we assume ideal redundancy and consider a single polarization mode to simplify the analysis. 

We simulate mock snapshot visibilities, $V^{i,j}(\nu)$, with $10s$ integration time for HIRAX-256 array with realistic baseline separation (i.e., $6.5m$ space along east-west and $8.5m$ along north-south) in a regular grid configuration. We assume a complete sky model and perfect redundancy throughout our simulation, i.e., all the antenna elements are treated identically, with ideal spacing and uniform beam responses. Therefore, the errors emanate only from the primary beam models used in visibility simulations. We use the Common Astronomy Software Application (CASA) \textit{simulator} tool \citep{TheCASATeam2022} for predicting visibilities when given a sky image $I(\boldsymbol{\hat\theta},\nu)$. The input sky image is the Global Sky Model (GSM) \citep{Zheng2016} map generated for HIRAX frequency range and telescope location using the PyGDSM package (see Figure \ref{fig:5}) along with the H\textsc{i} signal sky generated from a Gaussian random field output of the \textit{cora} package \citep{cora}. The mean temperature brightness of the H\textsc{i} sky image is assumed to take the form: 

\begin{figure}
\centering
    \includegraphics[width=8cm]{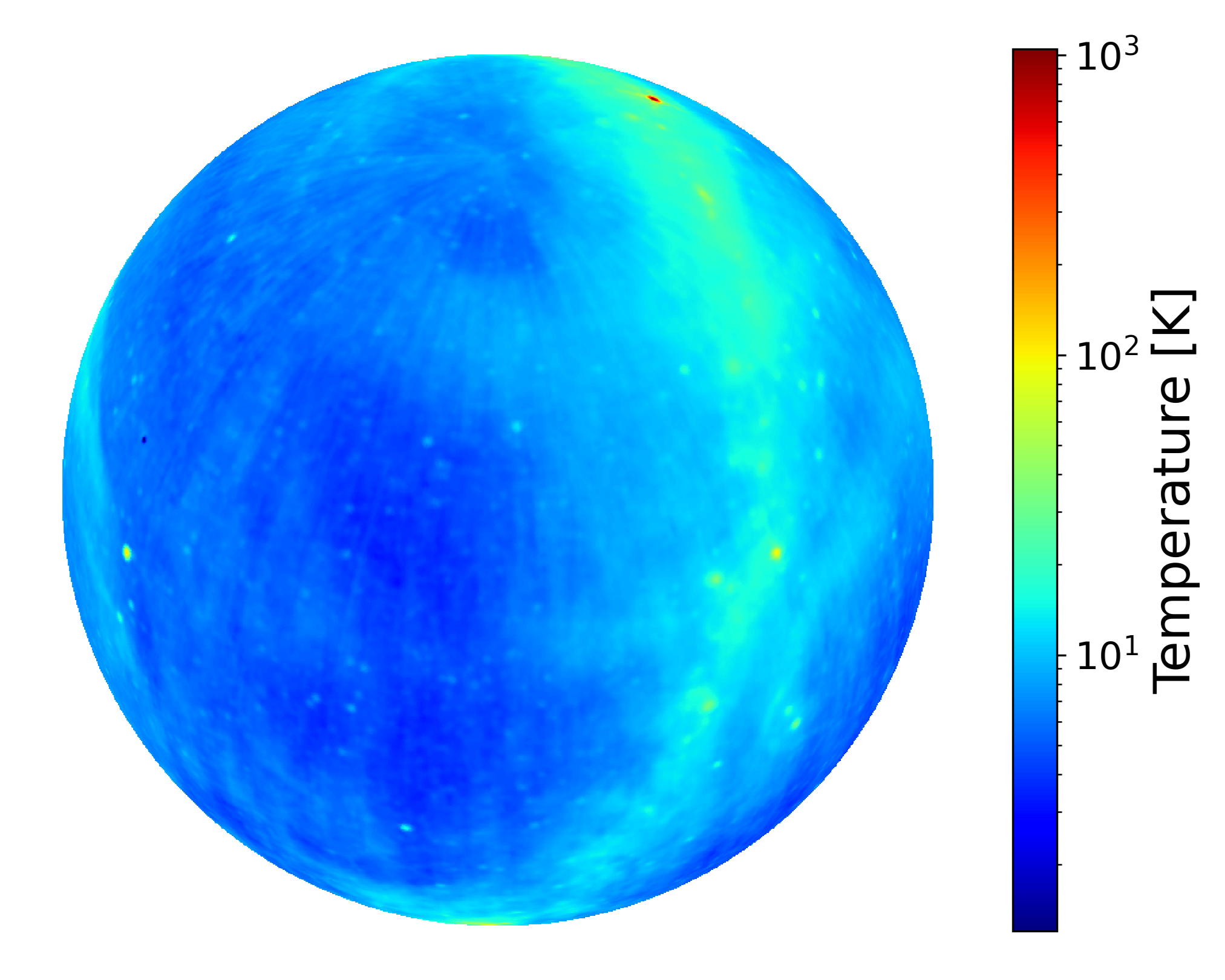}
    \caption{Chosen Global Sky Model at 600 $MHz$ generated at HIRAX location for the date 2024-09-05 at 00:00hrs UTC.\\}
    \label{fig:5}
\end{figure}

\begin{multline}
\label{eq:7}
 T_b(z) = 0.1 \left( \frac{\Omega_{H\textsc{i}}}{0.33 \times 10^{-4}} \right) \\ 
\times \left(\frac{\Omega_m + (1+z)^{-3} \Omega_{\Lambda}}{0.29}\right)^{-1/2} \left( \frac{1+z}{2.5}\right)^{1/2} ~mK .
\end{multline}

This 21cm sky model is taken from \citealp{Chang2008}, and \citealp{Shaw2015} and the cosmological parameters ($\Omega_{H\textsc{i}}, \Omega_m,$ and $ \Omega_{\Lambda}$) are taken from \citealp{planck2021}. The beam models are interpolated along the frequency direction to 1024 channels for more spectral resolution using spline interpolation. This interpolation is considered after a thorough investigation of the chromaticity of simulated CST beams, showing no drastic variations in the frequency structure beyond 5 $MHz$ frequency resolution.  

We simulate mock ``observed" visibilities $V^{i,j}_{obs}(\nu)$, 
\begin{multline}
\label{eq:8}
  V^{i,j}_{obs}(\nu) = \int \mathcal{B}_{true}(\boldsymbol{\hat\theta},\nu) [I_{GSM}(\boldsymbol{\hat\theta},\nu)+I_{H\textsc{i}}(\boldsymbol{\hat\theta},\nu)] \\
  \text{ exp } [-2\pi i \boldsymbol{u_{ij}}.\boldsymbol{\hat\theta}] ~ d\Omega
\end{multline}

 where $I_{GSM}(\boldsymbol{\hat\theta},\nu)$ is the global sky model, $I_{H\textsc{i}}(\boldsymbol{\hat\theta},\nu)$  is the H\textsc{i} 21cm signal sky distribution and $\mathcal{B}^{ij*}_{true}$ corresponds to the mock ``ground truth" beam. The ``model" visibilities $V^{i,j}_{mod}(\nu)$ are given by

 \begin{equation}
 \label{eq:9}
  V^{i,j}_{mod}(\nu) = \int \mathcal{B}_{mod}(\boldsymbol{\hat\theta},\nu) I_{GSM}(\boldsymbol{\hat\theta},\nu) \text{ exp } [-2\pi i \boldsymbol{u_{ij}}.\boldsymbol{\hat\theta}] ~ d\Omega
 \end{equation}
 
with the same sky model but different beam models $\mathcal{B}_{mod}$. Once the measurement sets are populated with observed visibilities and model visibilities, we perform model subtraction, i.e., subtract the ``model" visibilities (Equation \ref{eq:9}) from the ``observed" visibilities (Equation \ref{eq:8}) to recover H\textsc{i} visibilities. If we know the beam perfectly (perfect modeling), i.e., $\mathcal{B}_{true} = \mathcal{B}_{mod}$, then, $V^{i,j}_{rec}(\nu)$ reduces to H\textsc{i}-only visibility, 
\begin{multline}
\label{eq:10}
 V^{i,j}_{rec}(\nu) =  V^{i,j}_{H\textsc{i}}(\nu) \\
 =  \int \mathcal{B}_{true}(\boldsymbol{\hat\theta},\nu) I_{H\textsc{i}}(\boldsymbol{\hat\theta},\nu) \text{ exp } [-2\pi i \boldsymbol{u_{ij}}.\boldsymbol{\hat\theta}] ~ d\Omega .
\end{multline}

If $\mathcal{B}_{true} \neq \mathcal{B}_{mod}$, then there exists a beam modeling error term $\Delta\mathcal{B} = \mathcal{B}_{true} - \mathcal{B}_{mod}$, such that, 
\begin{multline}
\label{eq:11}
V^{i,j}_{rec}(\nu)=  \int \mathcal{B}_{true}(\boldsymbol{\hat\theta},\nu) I_{H\textsc{i}}(\boldsymbol{\hat\theta},\nu) \text{ exp } [-2\pi i \boldsymbol{u_{ij}}.\boldsymbol{\hat\theta}] ~ d\Omega \\ 
+ \int \Delta\mathcal{B}(\boldsymbol{\hat\theta},\nu) I_{GSM}(\boldsymbol{\hat\theta},\nu) \text{ exp } [-2\pi i \boldsymbol{u_{ij}}.\boldsymbol{\hat\theta}] ~ d\Omega .
\end{multline}

So, unmodelled foreground residuals are present in the recovered visibility, which can be removed up to some level with proper modeling of the primary beam. We will use our simulation setup here to study the sidelobes' mismodeling and non-modeling impacts in recovering H\textsc{i} signal power spectrum \footnote{This simulation setup is not the complete or the final HIRAX analysis pipeline, but rather a simple tool developed specifically for this work. A more robust and complete pipeline is in the works}.

\begin{table}
\begin{tabular}{c|c}
\hline
\hline
\multicolumn{2}{c}{CASA simulation parameters}\\
\hline
\hline
Sky image & GSM \\  
$N_{channels}$ & 1024 \\  
Channel width & 390.625 kHz \\  
Map size & $1501~\times~1501$ pix \\ 
Angular resolution & 0.1 deg \\  
$N_{antennas}$ & 256 \\  \hline 
\hline
\end{tabular}
\caption{ Parameters used for visibility simulation using the CASA simulator tool.}
\label{table2} 
\end{table}

We apply the Fourier transform to each baseline visibility along the frequency axis and square its amplitude. This squared quantity is related to the observed power spectrum in cosmological coordinates. The baseline ($\textbf{b}$), and delay (\textbf{${\eta}$}) axes are mapped to cosmological cylindrical Fourier modes $k_{\parallel}$ and $k_{\bot}$ to calculate the 2DPS given by: 

\begin{equation}
\label{eq:12}
  P(k_\bot, k_\parallel) \equiv \frac{c(1+z_0)^2 r^2(\chi(z_0))}{\nu_{21} H(z_0) W} \Delta\nu^2 |\text{ FFT}[\phi(\nu) V^{i,j}(\nu)]|^2
\end{equation}
where $c$ is the velocity of light in free space, $\nu_{21}$ is the rest frequency of the 21cm spin-flip transition of H\textsc{i}, $z_0$ is the corresponding redshift to the band center, W is the volume kernel introduced by the primary beam and the spectral window function $\phi(\nu)$, $r(\chi(z_0))$ is the comoving distance at $z_0$, $\Delta\nu$ is the channel width, and $H (z)$ is the Hubble parameter. The power spectrum formalism is explained in detail in Appendix \ref{sec:ps_form}. 

\subsection{Impact of beam modeling errors}\label{sec:impact}

Now that we have explained the simulation method, we change the beam model $\mathcal{B}_{mod}$ in Equation \ref{eq:9} to understand the effects of different beam models in power spectrum space. So, we look at different simulation configurations:  

\begin{itemize}
    \item \textbf{(a)}$\mathbf{ GSM+H\textsc{i} : }$ A mock observation as per Equation \ref{eq:8}. This is the ``observed" visibility that uses a ground truth beam (the best-fit model obtained from the beam fitting procedure as in Section \ref{sec:zfit}) and GSM+H\textsc{i} sky.
    
    \item \textbf{(b)}$\mathbf{ H\textsc{i}: }$ A fiducial H\textsc{i}-only visibility where the ground truth beam and foreground free sky are used. This will be used to compare the recovered visibilities in the latter cases.  
    
    \item \textbf{(c) True Beam : } A foreground-only (GSM) model visibility (as per Equation \ref{eq:9}), that when the perfect beam model is used, produces an ideally recovered H\textsc{i} visibility as represented in Equation \ref{eq:10}. This should be precisely equal to the above case \textbf{b}, apart from some numerical errors in the pipeline and baseline coverage gaps that could add some minor errors. 
    
    \item \textbf{(d) \& (e) Naive and Rippled Gaussian : } Recovered visibilities as per Equation \ref{eq:11} using two different mainlobe models: (d) a naive Gaussian model assuming a smooth frequency-dependent FWHM, $ 1.03 ~ \lambda/D_{eff}$ represented by the dashed line in Figure \ref{fig:1}, which ignores sidelobes and misses main lobe chromaticity; and (e) an elliptical Gaussian model fitted to the CST simulated beam, as described in Section \ref{sec:gfit}, which captures the main lobe chromaticity of the HIRAX beam accurately.

    \item \textbf{(f), (g) \& (h) $\mathbf{ZT~n_{max} = 40,~60~\&~80}$ : } Recovered visibilities as per Equation \ref{eq:11} using a radial Zernike Transform (ZT) beam model fitted to the CST simulated beam, as described in Section \ref{sec:zfit}, with the following restrictions on the maximum $n$-th order: (f) $n_{max}$=40, corresponding to an angular extent of $\sim 30^\circ$ from the beam center; (g) $n_{max}$=60, corresponding to $\sim 45^\circ$; and (h) $n_{max}$=80, corresponding to $\sim 60^\circ$, all with a similar number of $az$-modes. \footnote{The maximum $m$-th order is set by the maximum of $n$-th order being chosen. However, one can set the contribution of $az$-modes being used as we do in cases $i,j \& k$. Also, denoting these cases as first and second sidelobes would be simpler, but the number of sidelobes is not the same throughout the band; they change with frequency}.

    \item \textbf{(i), (j) \& (k) $\mathbf{ ZT~m_{max} = 0,~2~\&~4}$ :} Recovered visibilities as per Equation \ref{eq:11} using an azimuthal Zernike Transform (ZT) beam model fitted to the CST simulated beam, as described in Section \ref{sec:zfit}. The model has fixed $n_{max} = 100$ at the highest frequency, covering the full radial extent. We increase the complexity of the model in the azimuthal direction $\phi$ by choosing (i)$m=0$ (azimuthally symmetric modes), (j) $m_{max} = 2$ (up to quadrupolar modes), and (k) $m_{max} = 4$ (up to octupolar modes).
    
\end{itemize}

\begin{figure*}
\centering
    \includegraphics[width=18cm]{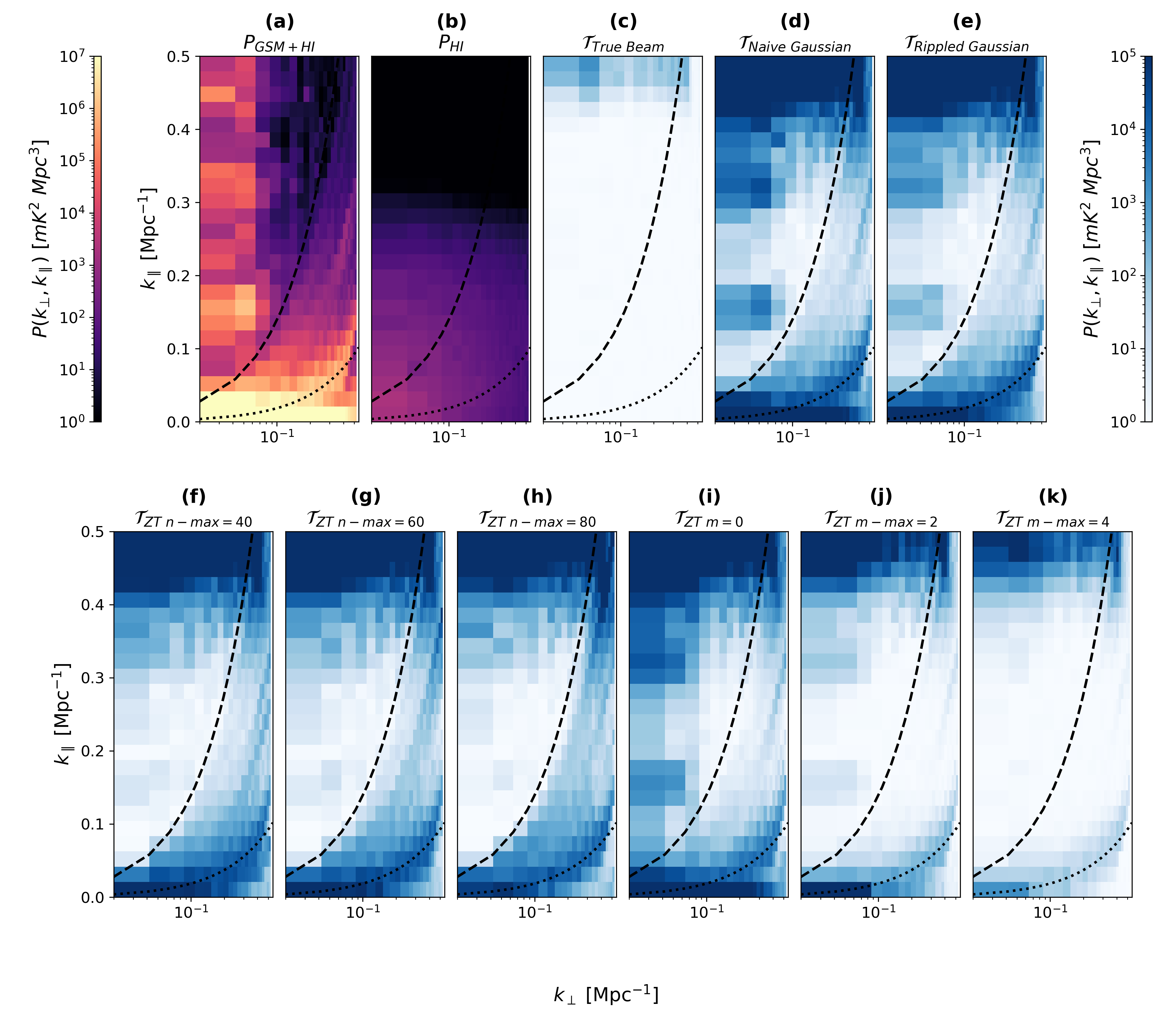}
    \caption{The 2-dimensional cylindrical power spectrum (2DPS) for the ``observed" visibility \textbf{(a)} and true H\textsc{i} visibility \textbf{(b)} and the 2D  transfer function $\mathcal{T}$ for all the other cases stated in Section \ref{sec:impact}. The amplitude is given in $mK^2 ~ Mpc^3$, and the dashed and dotted lines correspond to \textit{horizon} wedge ($\psi_0 = 90^\circ$) and \textit{primary field of view} wedge ($\psi_0 = 10^\circ$). See Appendix \ref{sec:ps_form} (Equation \ref{eq:b19}) for definition of wedge angle $\psi_0$. The $n$-cases (f,g \& h) has $n$-max = $m$-max, whereas, the $m$-cases (i,j \& k) has $n$-max = 100.  \\}
    \label{fig:6}
\end{figure*}    

\begin{figure*}
\centering
    \includegraphics[width=18cm]{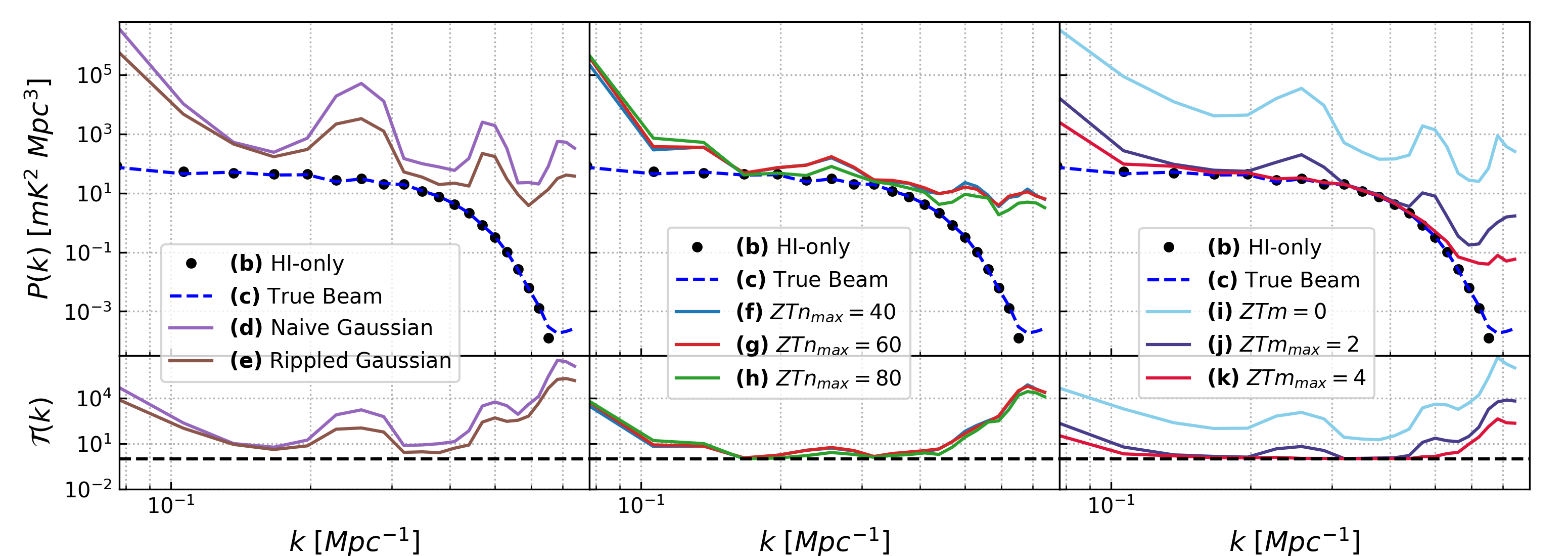}
    \caption{The 1-dimensional power spectrum (1DPS) (\textit{top}) and the transfer function ($\mathcal{T}$ - see Equation \ref{eq:14}) (\textit{bottom}) for all the cases stated in Section \ref{sec:impact}, with main lobe models, $rad$-mode cases and $az$-mode cases in separate plots from \textit{left} to \textit{right}. \\}
    \label{fig:7}
\end{figure*}
The 2-dimensional power spectrum (2DPS) is then computed from these visibilities (see Equation \ref{eq:12}). Since one may never be able to measure the true beam in reality to perfection, one needs to understand the effect of mismodeling or non-modeling certain features in the beam. To quantify these effects, we define a naive power spectrum transfer function ($\mathcal{T}$) \citep{Ansari2012, Masui2013, Anderson2018, Cunnington2022, Cunnington2023}, which is the ratio of recovered H\textsc{i} power spectrum to fiducial power spectrum in both 2D and 1D. We call this a naive transfer function because it is a bit different from the one used in \citealp{Cunnington2023}, where they use the ratio of cross-correlation between the input and the recovered signal to the autocorrelation of the input signal instead of the direct ratio of spectra. However, we stick to this naive transfer function for simplicity. This transfer function method has been proposed and tested to perform a correction to de-bias the recovered power spectrum to get closer to the true H\textsc{i} power spectrum. We define this naive transfer function as 
\begin{equation}
\label{eq:14}
    \mathcal{T}(k_\parallel,k_\bot) = \frac{P_{rec}(k_\parallel,k_\bot)}{P_{true}(k_\parallel,k_\bot)}.
\end{equation}
\change{In the ideal case of perfect foreground subtraction, $T(k)$ describes the purely multiplicative attenuation of the HI signal; in our simulations using a poor beam for the subtraction, $T(k)$ also implicitly includes an additive bias from residual foreground leakage specific to the chosen foreground realisation.}

Figure \ref{fig:6} shows the 2DPS for ``observed" and H\textsc{i}-only cases (a \& b) and 2-dimensional transfer function for all the other cases. The characteristic wedge feature can be seen in the foreground cases, where the most power resides. The black dashed and dotted lines denote the \textit{horizon} wedge and \textit{primary field of view} wedge, corresponding to an angle of $90^\circ$ and $10^\circ$, respectively. Any foreground power beyond the \textit{horizon} wedge is purely due to the chromaticity of the instrument. The ripple-like structure along the $k_\parallel$ direction at low $k_\bot$ is from the beam itself, manifesting from the solid angle of the beam (refer to appendix \ref{sec:solidang}). As the beam model approaches the true beam, the power in the recovered power spectrum progressively decreases. The clean signal window, i.e., white regions in the 2D transfer function plots, gets larger and cleaner as the beam model improves.  

We also compare all these cases through a 1D power spectrum (1DPS). The 2DPS can be radially averaged to get the 1DPS by averaging over $k$  by choosing a bin width $\Delta k$ ($\sim0.025 Mpc^{-1}$ in our case) radially, which is given by
\begin{equation}
\label{eq:13}
   P(k) = \frac{1}{N(k)} ~ \sum_i P(k_{\bot,i}, k_{\parallel,i}) \\               
\end{equation}
where $N(k)$ is the number of $(k_{\bot,i}, k_{\parallel,i})$ pairs within the bin centered at $k \equiv \sqrt{k_\bot^2 + k_\parallel^2} \in ~ \left[k - \frac{\Delta k}{2} ,k +  \frac{\Delta k}{2}\right]$. The wedge feature is masked so that we only take power outside the horizon limit for quantifying leakage (everything below the \textit{horizon} wedge). Figure \ref{fig:7} shows the 1DPS and the 1D naive transfer function (which can be calculated in the same way using Equation \ref{eq:13}) for both $rad$-mode and $az$-mode cases. 

By comparing the foreground residuals through the transfer function for different beam models, we can understand the significance of sidelobes in preserving the integrity of the H\textsc{i} signal. Also, the mainlobe-only models ($d~\&~e$) are insufficient since they still inherit the ripple structure. Additionally, the effects of missing out on the sidelobes seem entirely equivalent to assuming a symmetric model by visually comparing cases  $d$ and $i$, which strongly motivates us to include asymmetrical features in the model. In the $rad$-modes cases ($f,g~\&~h$), the leakage is lower than in the mainlobe-only case, but an excellent recovery is not entirely achievable without a perfect model. The ripple along the $k_\parallel$-axis in the low $k_\bot$ region is also reflected in 1DPS and the 1D transfer function. The specific scales at which these residuals pose serious challenges to H\textsc{i} recovery is around $k~\sim 0.25Mpc^{-1}$ and at $k~\sim 0.5Mpc^{-1}$ scales \footnote{We used the full HIRAX band for our power spectrum estimation centered at 600 $MHz$, to witness the effect of beam chromaticity of the instrument. So, our redshift bin is really wide, and thus, the $k$-values could be off by some fraction.}. These scales could be of higher significance and seem to be dominated by the characteristic ripple. This ripple should be addressed separately with utmost care in future works.  \change{In principle, the average frequency dependence of the beam solid angle (the characteristic ripple that we see in Figure \ref{fig:6}) and \ref{fig:7}) could be corrected using the factor defined in Equation \ref{eq:b21}. However, in practice, accurately measuring this solid angle quantity is highly challenging, particularly for individual antennas. On the other hand, one would need an accurate composite sky model or ideally be able to point to a field dominated by a bright point source in the beam centre during observation. This is difficult for HIRAX as it is a drift scan telescope. Therefore, these figures present the uncorrected beam, reflecting the realistic chromatic effects that would remain in the absence of precise beam measurements.} Some argue that the bias in these scales can be suppressed by adequately weighting the baselines \citep{Patil2016, EwallWice2017} and doing a proper bandpass calibration \citep{Seo2016}, but may not be entirely removed, and additional foreground filtering would be needed. This procedure has to be repeated with several foreground simulations to obtain the optimal transfer function required to recover an unbiased power spectrum. 

\section{Discussion}\label{sec:disc}
After demonstrating the importance of wide-field beam modeling, capturing azimuthal asymmetry, and characterizing the sidelobes and their chromaticity, several questions remain to be addressed. For example, how far into the sidelobes does one need to model to achieve high signal-to-noise ratio (SNR) detection of the cosmological H\textsc{i} signal? How many sidelobes can realistically be measured using techniques such as drone mapping, especially in the presence of system noise? The answers to these questions depend on several factors, including the fidelity of drone mapping, the quality of baseline redundancy, and the accuracy of the sky model. This work contributes to setting clear goals for drone mapping measurements, particularly regarding how far into the sidelobes we need to model to minimize foreground leakage.

This work aims to identify the effects of primary beam chromaticity in HIRAX beams and to set goals for drone calibration strategies. As previously established, there is a strong need for more $rad$-modes, as capturing up to a sizeable angular extent is essential to ensure no signal loss in this setup. However, one should also note that no foreground removal or systematic filtering methods were applied here. In contrast, extending to higher $az$-modes is not as crucial — capturing up to $m \leq 2$, or setting $m_{max}=2$, should be sufficient to avoid most issues in our case.  The follow-up work will address the issues with true beam maps measured through the drone mapping methods and apply this modeling technique. Also, we plan to propose and test mitigation strategies in the follow-up paper. Furthermore, future works will explore whether this approach holds for more complicated beam models, where closely packed dishes induce a change in beam shape, adding to the complexity. We have used Stokes I beam throughout the work, but cross-coupling effects will affect the voltage beams of individual antennas, and this problem has to be studied in the future. 

The primary hindrance in measuring the H\textsc{i} power spectrum with high precision is the chromaticity of the primary beam. To isolate the beam modeling bias, we have deliberately ignored several realistic error factors that are difficult to simulate. As a result, we assume noise-free visibilities, perfect redundancy, uniform beam responses across the array, and no beam squint. Additionally, we do not perform calibration or baseline weighting throughout this work. While this approach simplifies the analysis, it also introduces limitations. One potential shortcoming is the accurate modeling of the Stokes Q, U, and V beams, which we have not addressed in this study. Since our focus is on the Stokes I beam, we defer the issue of polarization leakage to future work. 

Even though the modeling errors introduce significant bias compared to the signal in the power spectrum, it is crucial to note that we omit an initial foreground filtering step.  Furthermore, any bias originating from modeling or calibration errors can be captured in delay-delay-rate space, where these systematics tend to be localized, and a temporal filter can then be applied to flag these errors from the analysis, as discussed in \citealp{Charles2023}. Previous studies have shown that principal component analysis (PCA) can clean foregrounds to a significant degree. More recently, specific machine learning techniques (especially U-Net-based algorithms) have been proposed to help signal recovery despite low-order foreground residuals and systematics \citep{Makinen2021, Ni2022}. Comparing the conventional filtering methods with the emerging machine learning-based approaches for foreground subtraction could open avenues for addressing persistent systematic biases in beam models, especially in cases where traditional approaches fall short. 

Finding an optimal transfer function has been proposed as a viable approach to de-bias the foreground-contaminated power spectrum. We provided a naive transfer function here based on the ratio of spectra alone. Hence, this method has to be tested rigorously before being used in actual analysis and repeated for many foregrounds and H\textsc{i} realizations to converge on an optimal transfer function for higher accuracy of signal reconstruction.

\section{Conclusion}\label{sec:conc}
In this work, we presented a robust beam modeling procedure that models the azimuthal asymmetries in the beam and isolates the sidelobes, enabling us to study their chromaticity separately. We emphasized the importance of modeling the mainlobe to the highest precision and the impact of sidelobes on the analysis in 21cm intensity mapping experiments. Using the power spectrum (2DPS and 1DPS) as a metric, we demonstrated the effect of incomplete beam modeling leading to the leakage of foregrounds in the 2D power spectrum. We also provide an idea of how this beam-related bias can be corrected in the future using a naive power spectrum transfer function ($\mathcal{T}(k)$). Incorporating the $az$-modes is a pivotal step forward in understanding the spatial features of primary beams of radio telescopes since the asymmetric features are usually neglected. The inclusion of asymmetric features in the basis stands out as a unique property of this modeling formalism, which enhances its importance and provides insights into the complex behavior of the beam. However, we also note that the discussed case of a few orders of asymmetries in the beam may get complicated if we consider feed legs, mutual coupling, and dish surface deformations, and this has to be tested in the future. 

We demonstrated the importance of modeling the chromatic sidelobes in this work, and we intend to investigate the extent of sky model accuracy where the beam knowledge is not complete in future works. Additionally, we plan to examine the effects of modeling the instrumental response when using machine learning foreground subtraction techniques in the future. This modeling approach will be tested on the accurate beam maps obtained from planned drone beam mapping flights in the HIRAX test-bed site. This work emphasizes the impact of sidelobes in 21cm power spectrum estimation and the role of the primary beam in the analysis pipeline. So, future analysis techniques for wide-field interferometers can benefit from including beam chromaticity as an essential factor that can contaminate the signal. Moreover, the setup used in this work is not the final or complete analysis pipeline for HIRAX, which is currently under development. This paper lays the groundwork for modeling and characterizing primary beam chromaticity. In a subsequent paper, we will extend this work to apply this method to a real drone-measured beam map and propose mitigation strategies.

\section*{Acknowledgments}
AS acknowledges the National Research Foundation (NRF) and The World Academy of Sciences (TWAS) Doctoral fellowship, which supported this work. DC acknowledges support from the  Swiss National Science Foundation grant 200021\_192243, and by SERI as part of the SKACH consortium. KM acknowledges support from the National Research Foundation of South Africa. HCC acknowledges the support of the Natural Sciences and Engineering Research Council of Canada (RGPIN-2019-04506, ALLRP 586202-23). EdLA acknowledges support from the UKRI STFC Rutherford Research Fellowship. QG acknowledges support from the UKRI Science and Technology Facilities Council (STFC) - grant number ST/W00206X/1. ERM acknowledges the support from the South African Radio Astronomy Observatory (SARAO). MGS acknowledges support from the South African Radio Astronomy Observatory and the National Research Foundation (Grant No. 84156).

\facilities{Computations were performed on Hippo at the University of KwaZulu-Natal, and the ilifu cloud computing facility - www.ilifu.ac.za, a partnership between the University of Cape Town, the University of the Western Cape, Stellenbosch University, Sol Plaatje University, and the Cape Peninsula University of Technology.} 

\software{The work made use of \texttt{numpy} \citep{Harris2020}, \texttt{scipy} \citep{Virtanen2020}, \texttt{matplotlib} \citep{Hunter2007}, \texttt{astropy} \citep{AstropyCollab} and \texttt{CASA} \citep{TheCASATeam2022}.\\}

\appendix
\section{Zernike transforms - Mathematical Formalism}\label{sec:zern}
The Wiener-Khinchin theorem states that the autocorrelation function is the Fourier transform of the power spectrum, which, in this context, implies that the far-field pattern of the antenna (i.e., beam) is closely related to the geometry of the aperture. To be more precise, the aperture illumination function is the Fourier transform of the far-field pattern. So, any initiative to model the beam response should consider the physical parameters of the antenna and the geometry of the aperture. To model the aperture space, one would innately choose ZP as the basis since they are orthogonal on a unit disk. Also, modeling the aperture illumination function with ZP is well established in the literature \citep{Sekhar2022}. The two significant advantages of choosing the ZP for this purpose are (i) they are orthogonal on a unit disk (dish aperture), and (ii) they can be factorized into a radial and a polar function as below: 
\\
\begin{equation}
\label{eq:a1}
Z^m_n(r,\psi) = R^m_n(r)e^{im\psi} \\  
\end{equation}

where $n,m \in \mathbb{Z}$ with $n\geq |m| \geq 0$, $n-m$ is even. The radial polynomial is defined as,
\begin{equation}
\label{eq:a2}
R^m_n(r) = \sum_{k=0}^{(n-m)/2} \frac{(-1)^k (n-k)!}{k! (\frac{n+m}{2} - k)! (\frac{n-m}{2} - k)!} r^{n-2k}. \\
\end{equation}

The angular functions are basic functions for two-dimensional rotation groups, and the radial functions are developed from well-known Jacobi polynomials \citep{BornWolf:1999:Book}). The ZP also satisfies the following orthogonal relationship on the unit circle \citep{Lakshminarayanan2011},
\begin{equation}
\label{eq:a3}
\int Z^m_n(r,\psi) ~ Z^{m'}_{n'}(r,\psi) ~ r ~ dr ~ d\psi = \frac{\varepsilon_m ~ \pi}{2n+2} ~ \delta_{nn'} ~ \delta_{mm'}
\end{equation}
\\
where $\varepsilon_m ~ = ~ 2$ if $m~=~0$, and 1 otherwise.

It is well known that a Fourier transform relates the aperture illumination and the beam response according to the van Cittert-Zernike theorem under the idealized assumption of no diffraction around the feed-can and feed-legs. So, we can use the sum of the Fourier transform of the ZP to describe the beam, for which we have a closed-form solution:

\begin{equation}
\label{eq:a4}
  \mathcal{Z}^m_n = z^m(\phi) ~ f_n (\theta) = \frac{e^{im\phi}}{2\pi i^m}
  ~ (-1)^{(n-m)/2} ~ \frac{J_{n+1}(\theta)}{\theta} 
\end{equation}
\\
where $\mathcal{Z}$ is factorized into angular and radial functions $z^m$ and $f_n$ and they are given explicitly given below 
\begin{equation}
\label{eq:a5}
z^m(\phi) = \frac{-1^{-m/2}}{2\pi i^m} e^{im\phi}
\end{equation} and 
\begin{equation}
\label{eq:a6}
f_n (\theta) = -1^{n/2} \frac{J_{n+1} (\theta)}{\theta} 
\end{equation} 
\\
where $J_{n+1}$ is the Bessel function of the first kind with order n+1. Asymptotically $J_{n+1}$ falls off as $1/\sqrt{\theta}$, therefore $f_n$ falls off as $\theta^{-3/2}$. 

This closed form of the Fourier transform of the ZP (Equation \ref{eq:1}) came from Nijboer-Zernike theory, which was used to represent the complex amplitude distribution of a point source image in the focal plane of an aberrated optical system \citep{nijboer}). 

We opt for Equation \change{\ref{eq:a4}} to be our basis function, but it is not orthogonal or complete, so we normalized the basis functions so that all functions have an equal area under the solid angle, using the relation,

\begin{equation}
\label{a7}
    \int_{-\infty}^{\infty} |A_n|^2 ~ |\mathcal{Z}_n^m \mathcal{Z}_n^{m*} |^2 ~ \theta ~ d\theta ~ d\phi~~= ~1 
\end{equation}

and we find the calculated normalization constant to be,
\begin{equation}
\label{eq:a8}
    A_n = \sqrt{\frac{(2n+1)(2n+3)(2n+5)}{(-1)^n}}.
\end{equation}

The number of modes for a given order $n$ for the chosen subset of the original basis is given by,

\begin{equation}
\label{eq:a9}
    N_{m \geq 0} = 
        \begin{cases}  
        \left(\frac{n+1}{2} \right)^2 & \text{for odd $n$},  \\ 
        \frac{n(n+2)}{4} & \text{for even $n$}.
        \end{cases}
\end{equation}

\begin{figure}
    \centering
    \includegraphics[width=5.8cm]{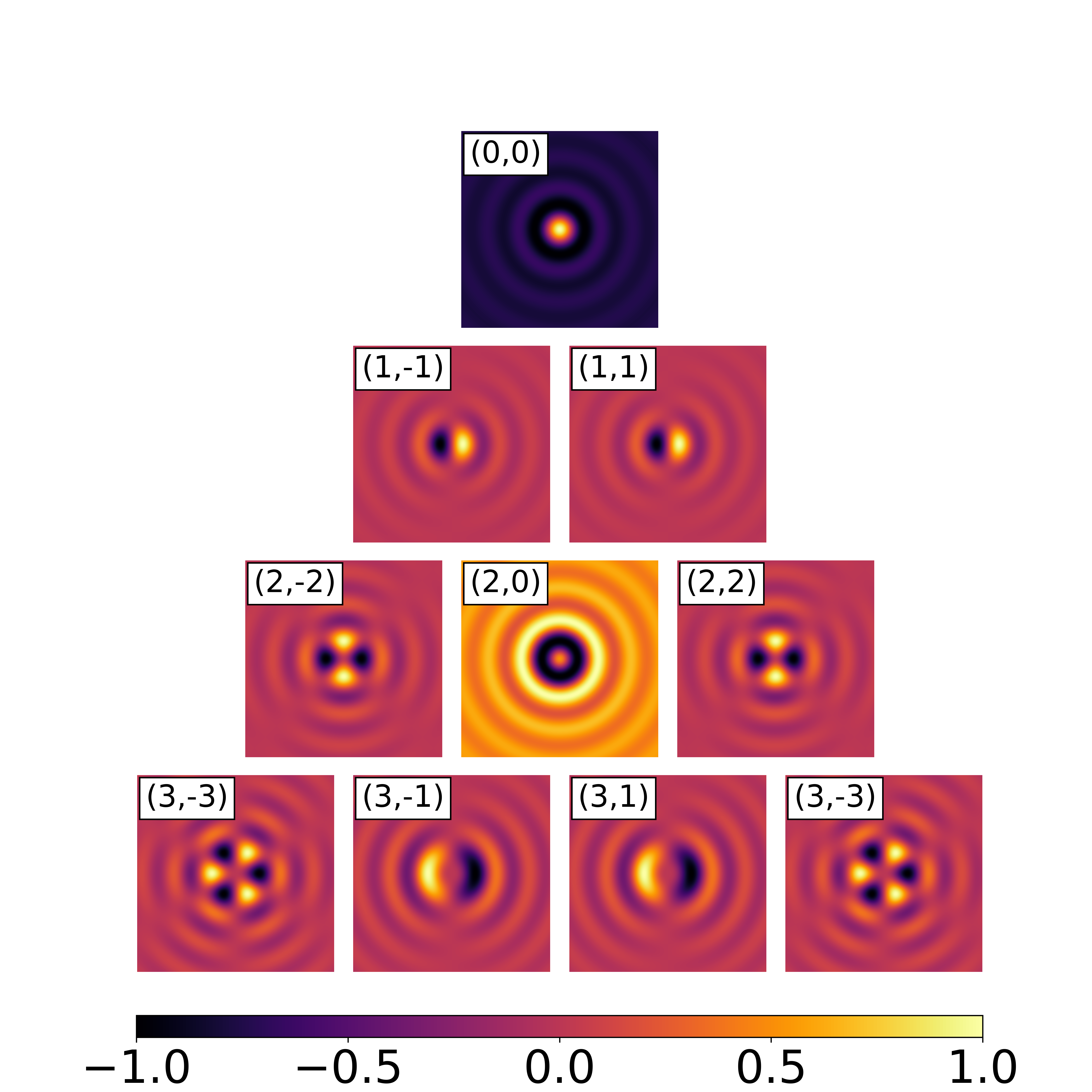}
    \quad
    \includegraphics[width=5.6cm]{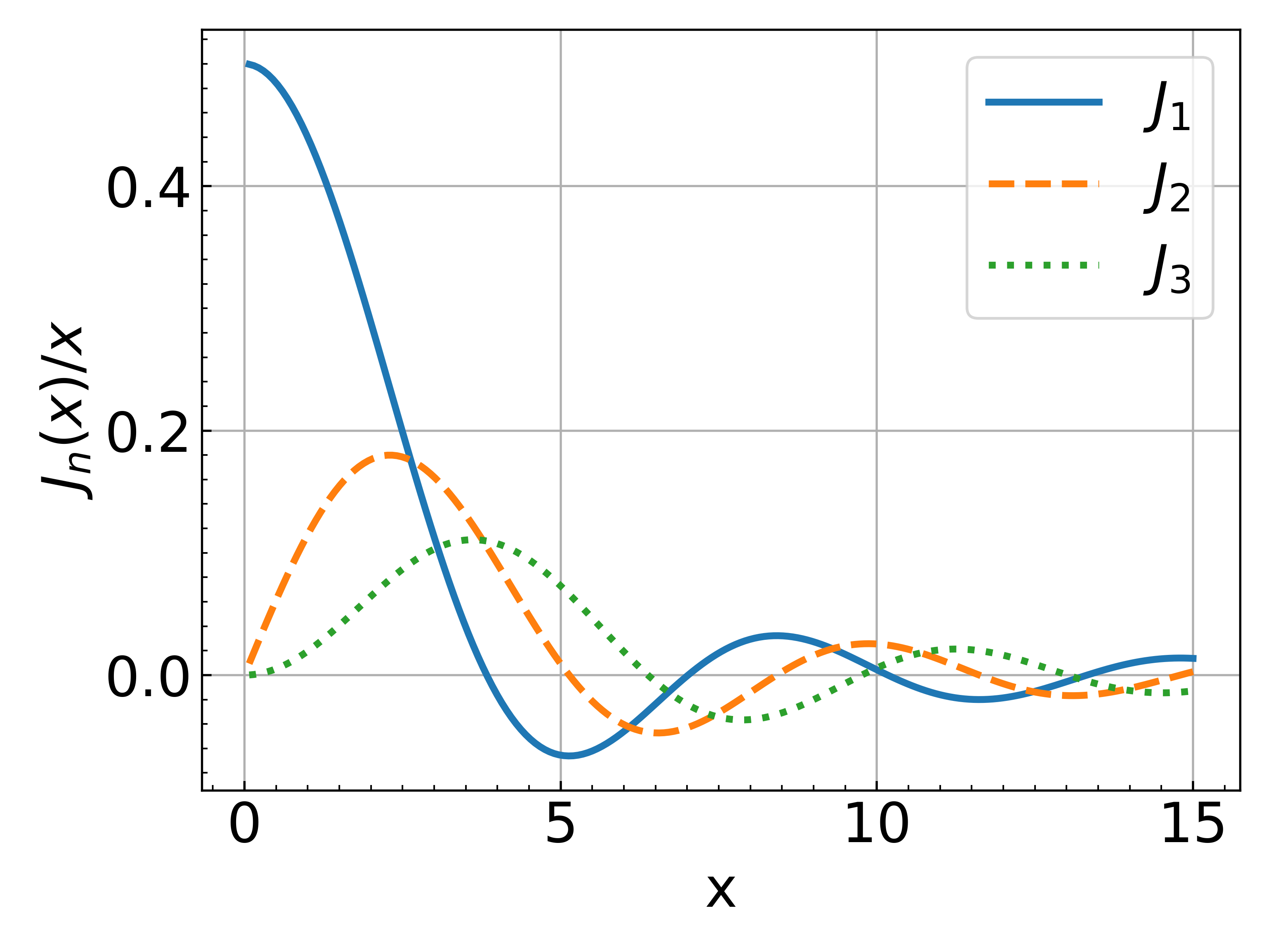}
    \quad
    \includegraphics[width=5.7cm]{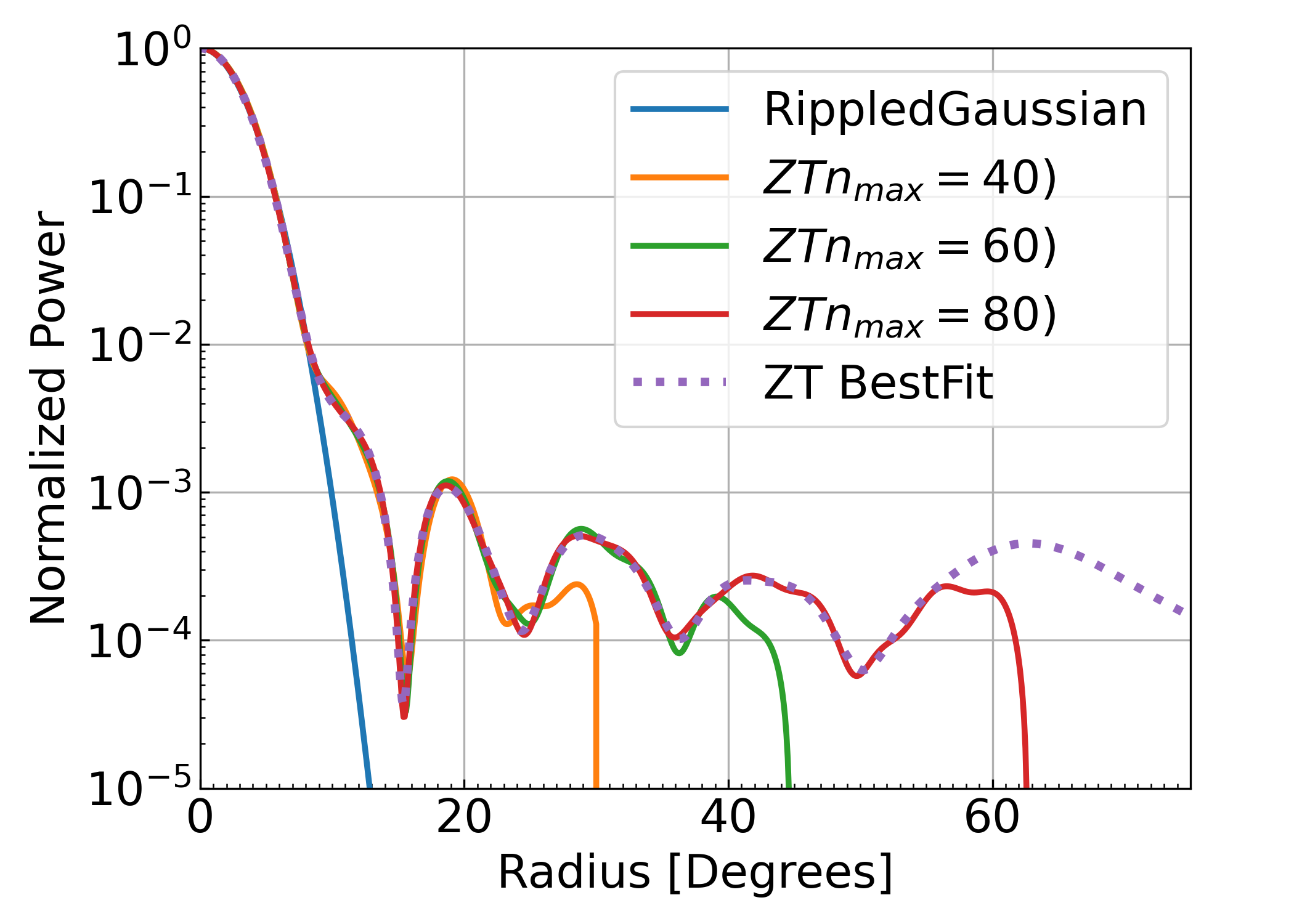}
    \caption{\textit{\textbf{Left:}} The real part of the analytical Fourier transform of the ZP. Since the real part of Equation \ref{eq:a6} is even for all values of $\textit{m}$, the negative and positive \textit{m}-modes are the same and create a degeneracy in the basis. So, to remove the degeneracy to make the basis more linearly independent, we choose a subset of the full basis, i.e., the real part of non-zero \textit{m}-modes. \textit{\textbf{Center:}} The shifting of scaled Bessel functions' maxima for the first 3 orders. The Bessel function scaled by the variable itself ($J_n(x)/x$) is plotted for increasing order of n, and it is clear that the higher-order functions have maxima farther from the center. \textit{\textbf{Right:}} The three considered ZT $rad$-mode models and their radial extent at 600 $MHz$. The ZT model does not have a sharp cutoff, as seen in this figure, but rather a small ringing tail, as seen in the central plot. We mask this tail for better visualization in logscale.\\}
    \label{fig:8}
\end{figure}

The radial polynomials (as in Equation \ref{eq:a6}) were previously used by \citealp{Hincks2010} to model the primary beam of the Atacama Cosmology Telescope, but they assumed an azimuthally symmetric beam profile.  This radial model (known as the Jinc function) corresponds to an unblocked circular aperture with uniform illumination \citep{Wilson2013}. Nevertheless, for HIRAX, we need to model the asymmetric structures as well, so we include the azimuthal (non-radial) modes, too. The higher $n$-order modes have their first maxima farther from the center (see Figure \ref{fig:1}), and the higher $m$-order modes have more azimuthally asymmetric features. We exploit this unique property of this basis to model the complexity of the HIRAX beam sidelobes. 

\section{Power spectrum formalism}\label{sec:ps_form}
Here, we briefly describe the formalism used to compute the power spectrum mentioned throughout this work, based on the per-baseline delay spectrum technique described in \citealp{Parsons2012}. The interferometric visibilities as measured by a radio interferometer \citep{vanCittert1934, Zernike1938} are given by the measurement equation,

\begin{equation}
\label{eq:b10}
    V^{i,j} (\nu) = \int \mathcal{B}_i \mathcal{B}_j^*(\boldsymbol{\hat\theta},\nu) ~ I(\boldsymbol{\hat\theta},\nu) ~ \text{ exp } [-2\pi i ~ \boldsymbol{u_{i-j}}.\boldsymbol{\hat\theta}] ~ d\Omega
\end{equation}

where $\boldsymbol{u_{i-j}} = \boldsymbol{b_{i-j}}/\lambda$ and $\boldsymbol{b_{i-j}}$ is the baseline vector joining two antennas $i$ and $j$; $\lambda$ is the wavelength at the band center; $\nu$ is the operating frequency; \textit{$\boldsymbol{\hat\theta}$} is the unit vector on the direction of the sky; \textit{c} is the speed of light, and \textit{$d\Omega$} is the solid angle element to which \textit{$\boldsymbol{\hat{\theta}}$} is normal. $\mathcal{B}_i \mathcal{B}_j^*$ reduces to $\mathcal{B}$ for a zero spacing baseline where $i=j$, i.e., $\mathcal{B} = |\mathcal{B}_i|^2$, and this denotes the antenna power pattern (beam term), and `\textit{I}' denotes the sky brightness. The Fourier transform of the visibilities along the frequency axis $\widetilde V^{i,j} (\eta)$, is, 

\begin{equation} 
\label{eq:b11}
    \widetilde V^{i,j} (\eta) = \int V^{i,j} (\nu) \phi(\nu) \text{ exp }[{-i(2\pi\nu\eta)}] d\nu 
\end{equation}

where $\eta$ is the delay and $\phi$ is the spectral weighting function, or more notably, the window function with which we can control the quality of the delay spectrum (for example, the Blackmann-Nutall window used in \citealp{Vedantham2012} and \citealp{Thyagarajan2013} and a self-convolved (squared in visibility space) Blackmann-Harris window used in \citealp{Thyagarajan2016}). And $\eta$ is the Fourier dual to $\nu$ and is directly related to the geometric signal delay between the antenna pairs, which is given by,
\begin{equation}
\label{eq:b12}
     \eta = \frac{\textbf{b} . \boldsymbol{\hat{\theta}}}{c}. 
\end{equation}
The Fourier relation in \ref{eq:b12} can be approximated in terms of a numerical Discrete Fourier Transform given by the Fast Fourier Transform (FFT) algorithm as,
\begin{equation}
\label{eq:b13}
     \widetilde V^{i,j} (\eta) = \Delta\nu \text{ FFT}[\phi(\nu) V^{i,j}(\nu)]   
\end{equation}

where $\phi(\nu)$ is a Blackmann-Harris window function. 
The square of $\widetilde V^{i,j} (\eta)$ in Equation \ref{eq:b14}, generally called the delay spectrum, is directly proportional to the observed spatial power spectrum,
\begin{equation}
\label{eq:b14}
    P_{obs}(\boldsymbol{u},\eta) \propto | \widetilde{V} (\boldsymbol{u},\eta) |^2 
\end{equation}
\begin{equation}
\label{eq:b15}
    P_{obs}(\boldsymbol{u},\eta) \equiv \frac{1}{W} ~ | \widetilde{V} (\boldsymbol{u},\eta) |^2 
\end{equation}

where W is the volume kernel introduced by the instrument and the choice of our window function, and is given by,

\begin{equation}
\label{eq:b16}
    W = \int_{-\infty}^{\infty} |\mathcal{B}_i \mathcal{B}_j^*(\boldsymbol{\hat\theta},\nu) ~ \phi(\nu)|^2 ~ d^2\boldsymbol{\hat\theta} ~ d\nu.
\end{equation}

The cosmological power spectrum in $k$-space at the band-centre is given by,

\begin{equation}
\label{eq:b17}
    P(k_\bot, k_\parallel) \equiv \frac{c(1+z)^2 r^2(\chi(z))}{\nu_{21} H(z)} P_{obs}(\boldsymbol{u},\eta) 
\end{equation}
 and,
\begin{equation}
\label{eq:b18}
    k_\bot = \frac{2 \pi \textbf{u}}{r(\chi(z))} ; \hspace{0.7cm} k_\parallel = \frac{2 \pi \nu_{21} H(z)}{c(1+z)^2} ~ \eta
\end{equation}

where $r(\chi(z))$ is the comoving distance along the line-of-sight direction, $\nu_{21}$ is the rest frequency of the 21cm spin-flip transition of H\textsc{i}, $z$ is the corresponding redshift to the band center, and $H (z)$ is the Hubble parameter and the foreground wedge is set by the relation, 

\begin{equation}
\label{eq:b19}    
    k_\parallel = k_\bot \frac{H(z)~r(\chi(z))}{c(1+z)^2} \psi_0
\end{equation}
where $\psi_0$ is the wedge angle ($\psi_0 = 90^\circ$ for horizon wedge). \\ 

And finally combining the relations \ref{eq:b13}, \ref{eq:b15}, \ref{eq:b17}, and \ref{eq:b18} , we get,
\begin{equation}
\label{eq:b20}
  P(k_\bot, k_\parallel) \equiv \frac{c(1+z_0)^2 r^2(\chi(z_0))}{\nu_{21} H(z_0) W} \Delta\nu^2 |\text{ FFT}[\phi(\nu) V^{i,j}(\nu)]|^2.
\end{equation}

\subsection{Ripple in Power spectrum}\label{sec:solidang}

\begin{figure*}
\centering
    \includegraphics[width=10cm]{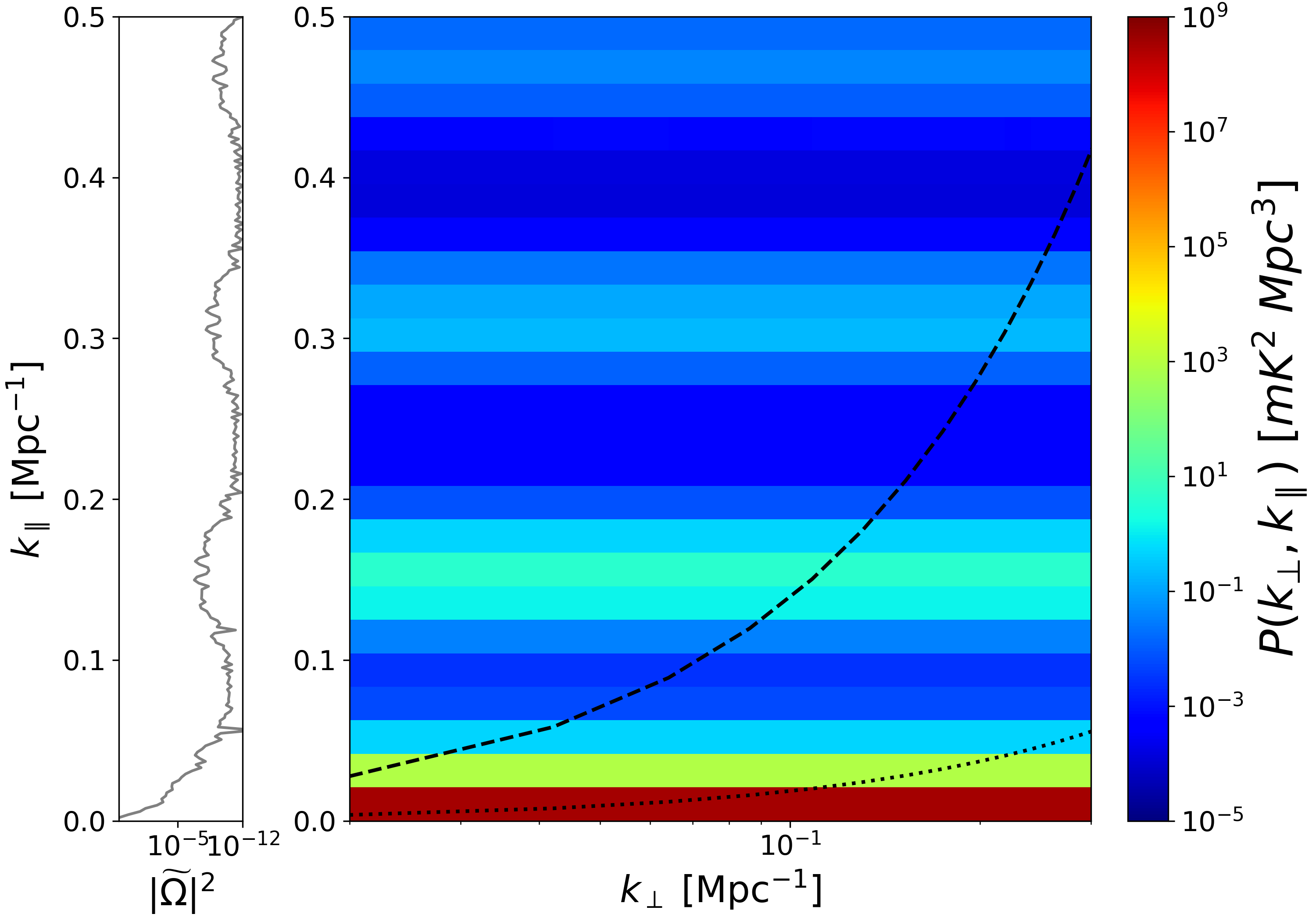}
    \caption{Source of ripple in 2D Power Spectrum. The calculated $\widetilde{\Omega}^2$ quantity as a function of $k_\parallel$ plotted against the 2DPS of a single point source using HIRAX CST beam.\\}
    \label{fig:9}
\end{figure*}

The inherent chromaticity of the HIRAX beam manifests as a ripple in power spectrum space along the $k_\parallel$ axis. We trace this back to the total power reception of the beam pattern at different frequencies, imparting a spectral structure. To visualize this in 2DPS, we simulate sample visibility with the same instrumental setup but chose a sky with a single point source at the field having no spectral structure (or flat in frequency, i.e., spectral index = 0) (refer to Figure \ref{fig:9}). \change{Also, in the limit of a uniform sky and zero spacing between antennas (autocorrelation), the Equation \ref{eq:b10} reduces to the beam solid angle, given by}
\begin{equation}
\label{eq:b21}
    \Omega_{ij}(\nu;\theta_{max}) = \int_{-\pi}^{\pi}\int_{0}^{\theta_{max}} \mathcal{B}_i(\theta,\phi,\nu)~\mathcal{B}_j^*(\theta,\phi,\nu) ~ sin\theta ~d\theta~d\phi
\end{equation}

\change{which for $i=j$ becomes $\Omega_{i} = \int |\mathcal{B}_i|^2~d\Omega$. We adopt $\theta_{max} = 75^\circ$ to match our map extent, making $\Omega(\nu)$ a truncated solid angle. This expression does not describe a point source (for which the angular integral collapses), but the zero-spacing response to the mean sky brightness.} We calculate the Fourier transform of this solid angle quantity $\Omega (\nu)$, and then square its amplitude and map it to cosmological $k_\parallel$ coordinates after weighting by a Blackmann-Harris window, using the relation from Equation \ref{eq:b18}. The calculated quantity is plotted alongside the 2DPS in Figure \ref{fig:9} and is given by 

\begin{equation}
\label{eq:b22}
    \widetilde{\Omega}(k_\parallel) =  \frac{2 \pi \nu_{21} H(z)}{c(1+z_0)^2} FFT[\phi(\nu) \Omega(\nu)].
\end{equation}

\bibliography{main}{}
\bibliographystyle{aasjournal}

\end{document}